\documentclass[PAPER, atlasdraft=false, USenglish, coverpage=false, cernpreprint=true, texmf, orcidlogo]{atlasdoc}
 
\usepackage[subfigure]{atlaspackage}
\usepackage{atlasbiblatex}
 
\usepackage{atlasphysics}
\usepackage{atlasjetetmiss}

\graphicspath{{logos/}{figures/}}
 
\usepackage{ANA-EXOT-2022-18-PAPER-defs}

\AtlasTitle{Pursuit of paired dijet resonances in the Run 2 dataset with ATLAS}

\AtlasAbstract{
New particles with large masses that decay into hadronically interacting particles are predicted by many models of physics beyond the Standard Model.
A search for a massive resonance that decays into pairs of dijet resonances is performed using 140~fb$^{-1}$ of proton--proton collisions at $\sqrt{s}=13$~TeV\ recorded by the ATLAS detector during Run~2 of the Large Hadron Collider.
Resonances are searched for in the invariant mass of the tetrajet system, and in the average invariant mass of the pair of dijet systems.
A data-driven background estimate is obtained by fitting the tetrajet and dijet invariant mass distributions with a four-parameter dijet function and a search for local excesses from resonant production of dijet pairs is performed.
No significant excess of events beyond the Standard Model expectation is observed, and upper limits are set on the production cross-sections of new physics scenarios.
}

\AtlasRefCode{EXOT-2022-18}
 
\PreprintIdNumber{CERN-EP-2023-147}

\HepDataRecord{140530}
 
\AtlasJournalRef{\PRD 108 (2023) 112005}
\AtlasDOI{10.1103/PhysRevD.108.112005}

\hypersetup{pdftitle={ATLAS document},pdfauthor={The ATLAS Collaboration}}
 
\begin{document}
 
\maketitle
 
\tableofcontents

\section{Introduction} 

New massive particles that decay into hadronically interacting quarks and gluons are predicted in many scenarios of physics beyond the Standard Model (BSM) accessible at the Large Hadron Collider (LHC), including well-motivated models of particle dark matter~\cite{Hinshaw:2012aka,Akrami:2018vks,Trimble1987,Bertone:2004pz,Feng:2010gw,Roos2012,Arrenberg:2013rzp,EXOT-2017-32} and models with large extra spatial dimensions~\cite{Boelaert:2010inn,Gingrich:2009hj,Calmet:2008dg,Giddings:2001bu,Dimopoulos:2001hw,Meade:2007sz,Anchordoqui:2003ug}.
Quarks and gluons produced at high energies fragment and hadronize into collimated \emph{jets} of particles~\cite{Salam:2010nqg}, observable in particle detectors like ATLAS~\cite{PERF-2007-01}.
The majority of Standard Model (SM) multijet event production occurs via \emph{nonresonant} quantum chromodynamics (QCD) processes, resulting in multijet systems with smoothly falling invariant mass distributions.
The large production cross-section of multijet processes can make searching for fully hadronic BSM signatures challenging, especially without the presence of other distinguishing feaures like leptons and/or missing transverse momentum~\cite{PERF-2016-07}.
However, when massive particles decay into pairs of jets (`dijets') via $s$-channel interactions, the invariant mass spectrum of the dijet system exhibits the signature of the massive particle as a \emph{resonance} around its mass value.
While the rate of new particle production may be too low that no resonance is obvious, such models may be detected using data-driven techniques analyzing the smoothly-falling invariant mass distribution of the SM background.
Searches for dijet resonances have been a cornerstone of collider physics at the LHC~\cite{EXOT-2010-01,CMS-EXO-10-010,EXOT-2010-07,CMS-EXO-11-015,EXOT-2011-21,EXOT-2011-07,CMS-EXO-11-094,CMS-EXO-12-016,EXOT-2013-11,CMS-EXO-12-059,CMS-EXO-14-005,EXOT-2015-02,CMS-EXO-15-001,CMS-EXO-16-032,EXOT-2016-21,CMS-EXO-16-056,EXOT-2019-03,CMS-EXO-19-012} and at earlier colliders~\cite{UA1:1988pgb,UA2:1993tee,CDF:2008ieg,D0:2009rxw}.
 
\begin{figure}[hbtp]
\centering
\includegraphics[width=0.5\textwidth]{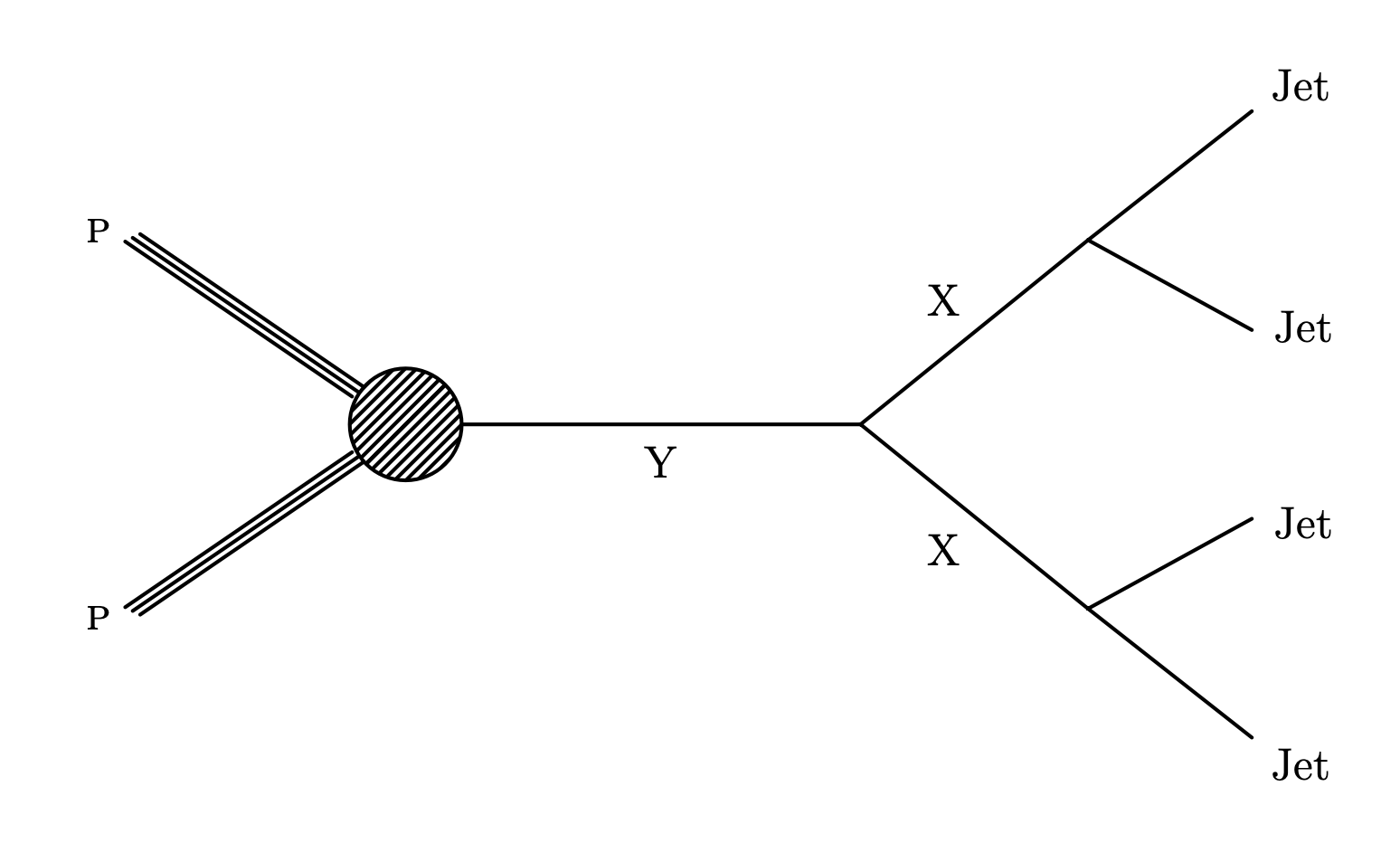}
\caption{A schematic representation of the signal topology studied in this analysis: a massive new particle $Y$ decays into two new particles with intermediate mass $X$, each decaying into a dijet system.}\label{fig:intro:cartoon}
\end{figure}

This paper presents a search for a generic massive resonance $Y$ that decays into two pairs of intermediate resonances $X$ with the same mass, each decaying into two partons and so typically producing a pair of dijet systems.
This decay structure is represented schematically in Figure~\ref{fig:intro:cartoon}.
Examples of exotic physics models that could produce such a final state topology include scalar diquark~\cite{Mohapatra:2007af,Han:2009ya,Richardson:2011df,Assad:2017iib} and coloron states~\cite{Chivukula:1996yr,Bai:2010dj,Chivukula:2013xka,Bai:2018jsr}, and additional new particle content such as a vector-like quarks that interact in pairs with the massive diquark or coloron and decay hadronically~\cite{Kilic:2008pm,Bai:2011mr,Dobrescu:2018psr}.
The analysis is performed on the Run 2 proton--proton ($pp$) collision data recorded by the ATLAS experiment.
Previous searches for signals with this resonance structure have been performed by the ATLAS~\cite{SUSY-2012-18,SUSY-2015-05,SUSY-2016-09,HDBS-2018-59} and CMS~\cite{CMS-EXO-12-052,CMS-EXO-17-021,CMS-EXO-21-010} collaborations at the LHC.
Most recently, in Ref.~\cite{CMS-EXO-21-010}, the CMS Collaboration studied final states where pairs of dijet resonances are collimated but insufficiently boosted to be reconstructed in a single large-radius jet (\emph{e.g.} a jet reconstructed with radius parameter $R$=0.8-1.0~\cite{CMS-JME-13-004,JETM-2018-06}).
A small, locally significant excess of events (3.9 standard deviations from two events, corresponding to a 1.6 standard deviation global significance) was observed with tetrajet resonance masses around $\mfour\sim8~\TeV$ producing dijet resonances with average masses around $\mtwo\sim2~\TeV$.
This prompted an investigation of such final state configurations using ATLAS data.
 
As there are two resonances with different masses involved in the final state topology ($Y$ and $X$), both the tetrajet system and the average dijet system invariant masses are separately studied using the \textsc{BumpHunter} algorithm~\cite{Choudalakis:2008pr,Aaltonen:2008vt,Choudalakis:2011qn,Vaslin:2022bds,pyBumpHunter}.
A data-driven background estimate is obtained by fitting these invariant mass distributions with a functional form.
 
An outline of the remainder of this paper is as follows.
Section~\ref{sec:atlas} provides overviews of the ATLAS detector, the Run 2 $pp$ data sample and the signal and background Monte Carlo (MC) simulations used in this search.
This is followed in Section~\ref{sec:methodology} by a description of the analysis methodology including jet reconstruction, event selection, the data-driven background estimation procedure and the systematic uncertainties considered.
The main results are presented in Section~\ref{sec:results}, interpreting the observed data in terms of upper limits on the production cross-sections of new physics scenarios.
Concluding remarks are made in Section~\ref{sec:conclusions}.


\section{ATLAS, the Run 2 data, and simulation}
\label{sec:atlas}

\newcommand{\AtlasCoordFootnote}{
ATLAS uses a right-handed coordinate system with its origin at the nominal interaction point (IP) in the center of the detector and the \(z\)-axis along the beam pipe. The \(x\)-axis points from the IP to the center of the LHC ring, and the \(y\)-axis points upwards. Cylindrical coordinates \((r,\phi)\) are used in the transverse plane, \(\phi\) being the azimuthal angle around the \(z\)-axis. The pseudorapidity is defined in terms of the polar angle \(\theta\) as \(\eta = -\ln \tan(\theta/2)\). Angular distance is measured in units of \(\Delta R \equiv \sqrt{(\Delta y)^{2} + (\Delta\phi)^{2}}\), where $y=(1/2)[(E+p_z)/(E-p_z)]$ is the object's rapidity defined by its energy and longitudinal momentum.}

\subsection{The ATLAS detector} 
 
The ATLAS detector~\cite{PERF-2007-01} at the LHC covers nearly the entire solid angle around the collision point.\footnote{\AtlasCoordFootnote}
It consists of an inner tracking detector surrounded by a thin superconducting solenoid, electromagnetic and hadron calorimeters,
and a muon spectrometer incorporating three large superconducting air-core toroidal magnets.
 
The inner-detector system is immersed in a \SI{2}{\tesla} axial magnetic field
and provides charged-particle tracking in the range of \(|\eta| < 2.5\).
The high-granularity silicon pixel detector covers the vertex region and typically provides four measurements per track,
the first hit normally being in the insertable B-layer 
installed before \RunTwo~\cite{ATLAS-TDR-19,PIX-2018-001}.
It is followed by the silicon microstrip tracker
, which usually provides eight measurements per track.
These silicon detectors are complemented by the transition radiation tracker (TRT),
which enables radially extended track reconstruction up to \(|\eta| = 2.0\).
The TRT also provides electron identification information
based on the fraction of hits (typically 30 in total) above a higher energy-deposit threshold corresponding to transition radiation.
 
The calorimeter system covers the pseudorapidity range of \(|\eta| < 4.9\).
Within the region \(|\eta|< 3.2\), electromagnetic calorimetry is provided by barrel and
endcap high-granularity lead/liquid-argon (LAr) calorimeters,
with an additional thin LAr presampler covering \(|\eta| < 1.8\)
to correct for energy loss in material upstream of the calorimeters.
Hadron calorimetry is provided by the steel/scintillator-tile calorimeter,
segmented into three barrel structures with \(|\eta| < 1.7\), and two copper/LAr hadron endcap calorimeters.
The solid angle coverage is completed with forward copper/LAr and tungsten/LAr calorimeter modules
optimized for electromagnetic and hadronic energy measurements respectively.
 
The muon spectrometer 
comprises separate trigger and
high-precision tracking chambers measuring the deflection of muons in a magnetic field generated by the superconducting air-core toroidal magnets.
The field integral of the toroids ranges between \num{2.0} and \SI{6.0}{\tesla\metre}
across most of the detector.
Three layers of precision chambers, each consisting of layers of monitored drift tubes, cover the region \(|\eta| < 2.7\),
complemented by cathode-strip chambers in the forward region, where the background is highest.
The muon trigger system covers the range of \(|\eta| < 2.4\) with resistive-plate chambers in the barrel, and thin-gap chambers in the endcap regions.
 
Interesting events are selected by the first-level trigger system implemented in custom hardware,
followed by selections made by algorithms implemented in software in the high-level trigger~\cite{TRIG-2016-01}.
The first-level trigger accepts events from the \SI{40}{\MHz} bunch crossings at a rate below \SI{100}{\kHz},
which the high-level trigger further reduces to record events to disk at about \SI{1}{\kHz}.
 
An extensive software suite~\cite{ATL-SOFT-PUB-2021-001} is used in data simulation, in the reconstruction and analysis of real
and simulated data, in detector operations, and in the trigger and data acquisition systems of the experiment.
 
\subsection{The Run 2 data sample} 
 
This analysis is performed using data from LHC $pp$ collisions with $\sqrt{s}=13$~\TeV, collected during 2015--2018 with the ATLAS detector.
The total integrated luminosity of this data sample is $140~\ifb$.
The uncertainty in the combined 2015--2018 integrated luminosity is 0.83\%~\cite{DAPR-2021-01}, obtained using the LUCID-2 detector~\cite{LUCID2} for the primary luminosity measurements.
Due to the high instantaneous luminosity and the large total inelastic $pp$ cross-section, there are, on average, 33.7 simultaneous  collisions (`\pileup') in each bunch crossing.
Data are required to satisfy certain quality requirements~\cite{DAPR-2018-01} to be included in the analysis.
 
During certain data-taking periods, modules of the tile calorimeter were disabled.
A study of the impact of vetoing these disabled modules in MC and data was performed, and found to have a negligible impact on the background shape modeling and expected limits.
 
\subsection{Simulated event samples} \label{sec:signal}
 
Samples of MC simulated signal and background (multijet) events are used for optimisation, estimation of possible signal contributions, and validation of background estimation strategies.
 
\PYTHIA~8.230~\cite{Sjostrand:2014zea,Sjostrand:2006za} was used as the nominal MC generator for both the signal and the background events.
Events were simulated using the A14 set of tuned parameters (`tune')~\cite{ATL-PHYS-PUB-2014-021}, the Lund string hadronization model and the \NNPDF[2.3lo]~\cite{Ball:2012cx} leading-order (LO) parton distribution function (PDF) set.
The \PYTHIA parton shower algorithm uses a dipole-style transverse momentum (\pt) ordered evolution, and its renormalization and factorization scales were set to the geometric mean of the squared transverse masses of the outgoing particles.
EvtGen~\cite{Lange:2001uf} was used to model decays of heavy flavor hadrons.
 
Signal samples were generated with \PYTHIA~8.230 using the process $W' \rightarrow WZ$, where the mass of the $W'$ corresponds to the $Y$ mass, and the $W/Z$ masses were both set to be equal to the $X$ mass.
The $W$ and $Z$ full widths at half maximum of a relativistic Breit--Wigner were set to 0.1~\GeV, so that the width of the resonance is determined by the detector resolution (typically ranging between 1-4\% for both \mfour~and~\mtwoavg).
These exotic $X$ bosons were forced to decay into quark--antiquark pairs, and decays into top--antitop quark pairs were disabled.
Representative ~\mfour~ and ~\mtwoavg~ distributions are shown in Figure~\ref{fig:signalshape} for several different choices of $\alpha=\mtwoavg/\mfour$ (see Section~\ref{sec:method:evsel}) with $m_{Y} = 6000$~\GeV.
The signal distributions have clear peaks near the generated signal masses.
 
\begin{figure}[htbp]
\centering
\subfigure[]{\includegraphics[width=0.48\textwidth]{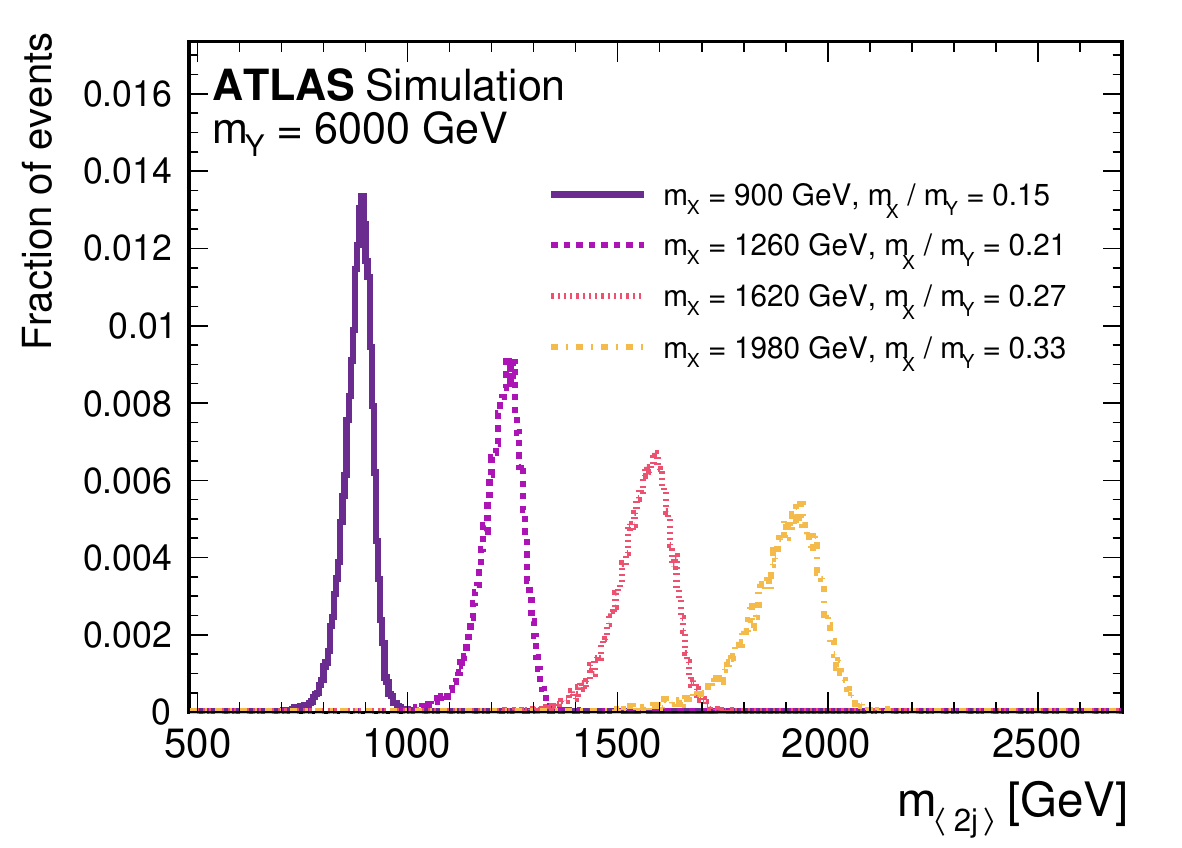}}
\subfigure[]{\includegraphics[width=0.48\textwidth]{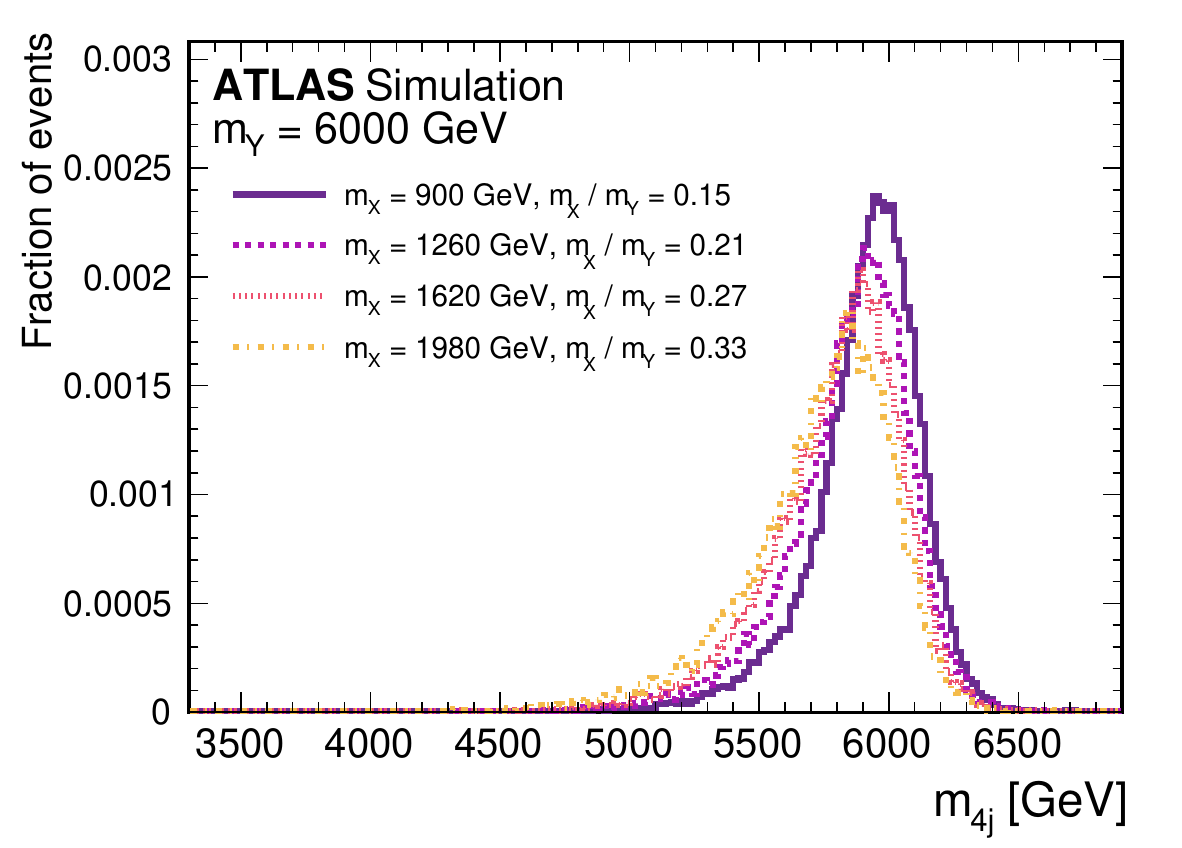}}
\caption{Examples of (a) ~\mtwoavg\ and (b) \mfour\ ~distributions for $m_{Y} = 6000$~\GeV with $m_{X}/m_{Y}=0.15$, $0.21$, $0.27$, and $0.33$.}
\label{fig:signalshape}
\end{figure}

Background samples of `hard-QCD' multijet processes were simulated using the same \PYTHIA settings.
These samples were used to optimize aspects of the analysis in early stages, although they are not used for the final background estimate.
Additional background multijet samples were simulated with \SHERPA[2.2.5]~\cite{Gleisberg:2008ta} to test the robustness of the background estimation procedure, using the default AHADIC cluster hadronization model~\cite{Webber:1983if}.
This sample includes LO matrix element calculations for \TwoToTwo processes, and used the \SHERPA parton shower algorithm based on Catani--Seymour dipole subtraction~\cite{Schumann:2007mg}.
It used the \CT[14nnlo] next-to-next-to-leading-order (NNLO) PDF~\cite{Dulat:2015mca} set for matrix element calculations and \CT[10] for multi-parton interactions~\cite{Lai:2010vv}.
 
Simulated background events were passed through a detailed detector simulation~\cite{SOFT-2010-01} based on \GEANT~\cite{Agostinelli:2002hh}, while simulated signal events were reconstructed with a fast simulation that uses a parameterisation of the ATLAS calorimeter response~\cite{ATL-PHYS-PUB-2010-013}.
In both cases, the samples were overlayed with minimum-bias interactions simulated using \PYTHIA[8]~with the A3 tune~\cite{ATL-PHYS-PUB-2016-017} and the \NNPDF[2.3lo] PDF set to represent \pileup interactions.
The distribution of the average number of \pileup interactions in simulation is reweighted during data analysis to match that observed in \RunTwo data.
 
Additional details of the MC samples used in this measurement may be found in Ref.~\cite{ATL-PHYS-PUB-2019-017}.


\section{Methodology}
\label{sec:methodology}

\subsection{Particle flow jets}
 
Jets are reconstructed from particle flow objects~\cite{PERF-2015-09} using the \akt algorithm~\cite{Cacciari:2008gp} as implemented in \Fastjet~\cite{Fastjet}, using a jet radius parameter $R=0.4$.
The ATLAS particle flow algorithm combines measurements from the ATLAS inner detector and calorimeter systems~\cite{PERF-2014-07} to improve the jet energy resolution (JER), reduce sensitivity to \pileup effects, and improve the jet reconstruction efficiency (especially at low jet \pt) relative to the jet reconstruction based on calorimeter signals alone.
Jets are required to have a $\pt > 60$~\GeV~and a rapidity $y$ satisfying $|y|<2.4$.
The jet energy scale (JES) of particle flow jets is calibrated using a combination of simulation-based and \emph{in situ} corrections~\cite{JETM-2018-05}.
 
\FloatBarrier
 
\subsection{Event selection}\label{sec:method:evsel}
 
To be considered for analysis, all detector-level events are required to have at least one primary vertex reconstructed from two or more inner-detector tracks with $\pt > 500$~\MeV.
Events are required to have at least four jets, from which two dijet pairs are reconstructed.
In the selected events, an event selection similar to that of Ref.~\cite{CMS-EXO-21-010} is applied to allow direct comparisons of the two searches.
The two dijet pairs are determined by minimizing the $\Delta R$ between the jets, defined as
\begin{equation*}
\Delta R = | \Delta R_{\text{AB}} - 0.8 | + | \Delta R_{\text{CD}} - 0.8 |,
\end{equation*}
where A, B, C, and D are the ordering of the four highest-\pt jets in the event that minimizes $\Delta R$.
The value of 0.8 ensures that the reconstructed pair of dijet systems are collimated, but not so boosted that they will be reconstructed as a \largeR jet.
Once the two dijet pairs AB and CD are selected, the dijet systems are required to satisfy angular requirements
\begin{equation*}
\Delta R_{\text{AB}} < 2.0, \; \Delta R_{\text{CD}} < 2.0
\end{equation*}
and
\begin{equation*}
\Delta \eta = |\eta_{\text{AB}} - \eta_{\text{CD}}| < 1.1.
\end{equation*}
In addition, the mass asymmetry between the two dijet pairs is required to satisfy
\begin{equation*}
\frac{m_{\text{AB}} - m_{\text{CD}}}{m_{\text{AB}} + m_{\text{CD}}} < 0.1.
\end{equation*}
 
After the event selection procedure is complete, the observables of interest are the invariant mass of the tetrajet system,~\mfour~(a proxy for $m_{Y}$), and the average invariant mass of the two dijet systems,~\mtwoavg~(a proxy for $m_{X}$).
The ratio of these quantities, $\alpha=\mtwoavg/\mfour$, is used to parameterize the kinematic space studied in this search in terms of the Lorentz boost of the $X$ decay products.
The correlations between~\mfour~and~\mtwoavg~are shown in Figures~\ref{fig:evsel:m4j:m2javg:data}~and~\ref{fig:evsel:m4j:m2javg:sig} after the event selection for the Run 2 data and for a simulated signal sample with $m_{Y}=6~\TeV$ and $m_{X}=2~\TeV$, respectively.
As shown, requirements on~\mtwoavg~are correlated with~\mfour\ and therefore they can sculpt the background mass distribution.
Figures~\ref{fig:evsel:m4j:alpha:data}--\ref{fig:evsel:m2javg:alpha:sig} illustrate the correlation between~\mfour,~\mtwoavg~and~$\alpha$.
For the background distribution, $\alpha$ is less correlated with~\mfour~than~\mtwoavg.
The analysis is performed in regions of~$\alpha$~rather than~$\mtwoavg$, to reduce background sculpting in the tetrajet and average dijet invariant mass spectra due to the selection criteria.
Twelve different $\alpha$ regions are used, evenly spaced to cover $0.10 < \alpha < 0.34$.
For each $\alpha$ region, separate fits are performed for~\mfour~and~\mtwoavg.
 
The combined acceptance times efficiency of all analysis selections is between 12\%--45\% for signal events as a function of the signal particle masses $m_{Y}$ and $m_{X}$
for $2000~\GeV < m_{Y} < 10000$~\GeV, and $500~\GeV < m_{X} < 3300$~\GeV.
 
\begin{figure}[htbp]
\centering
\subfigure[]{\includegraphics[width=0.4\textwidth]{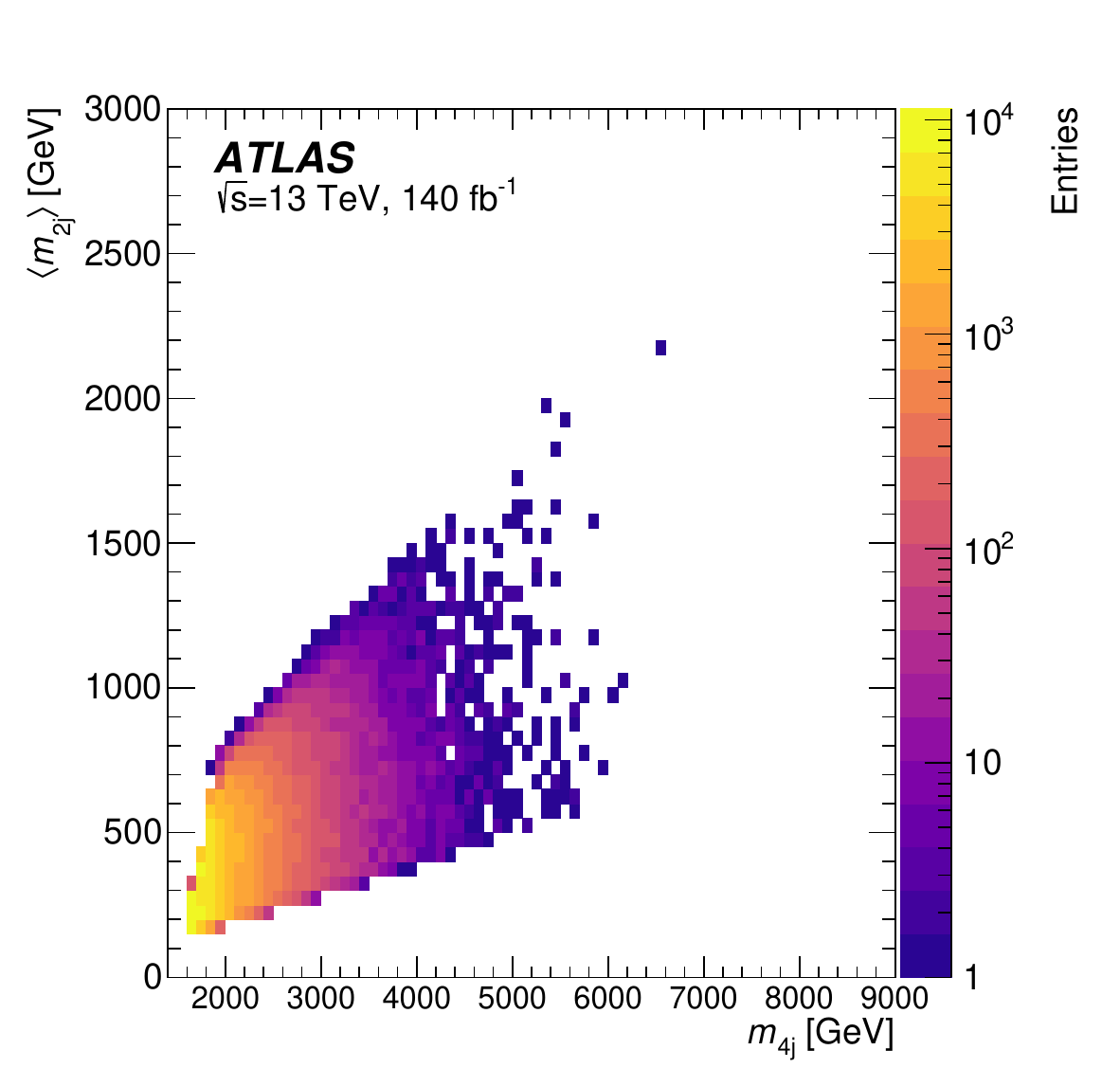}\label{fig:evsel:m4j:m2javg:data}}
\subfigure[]{\includegraphics[width=0.4\textwidth]{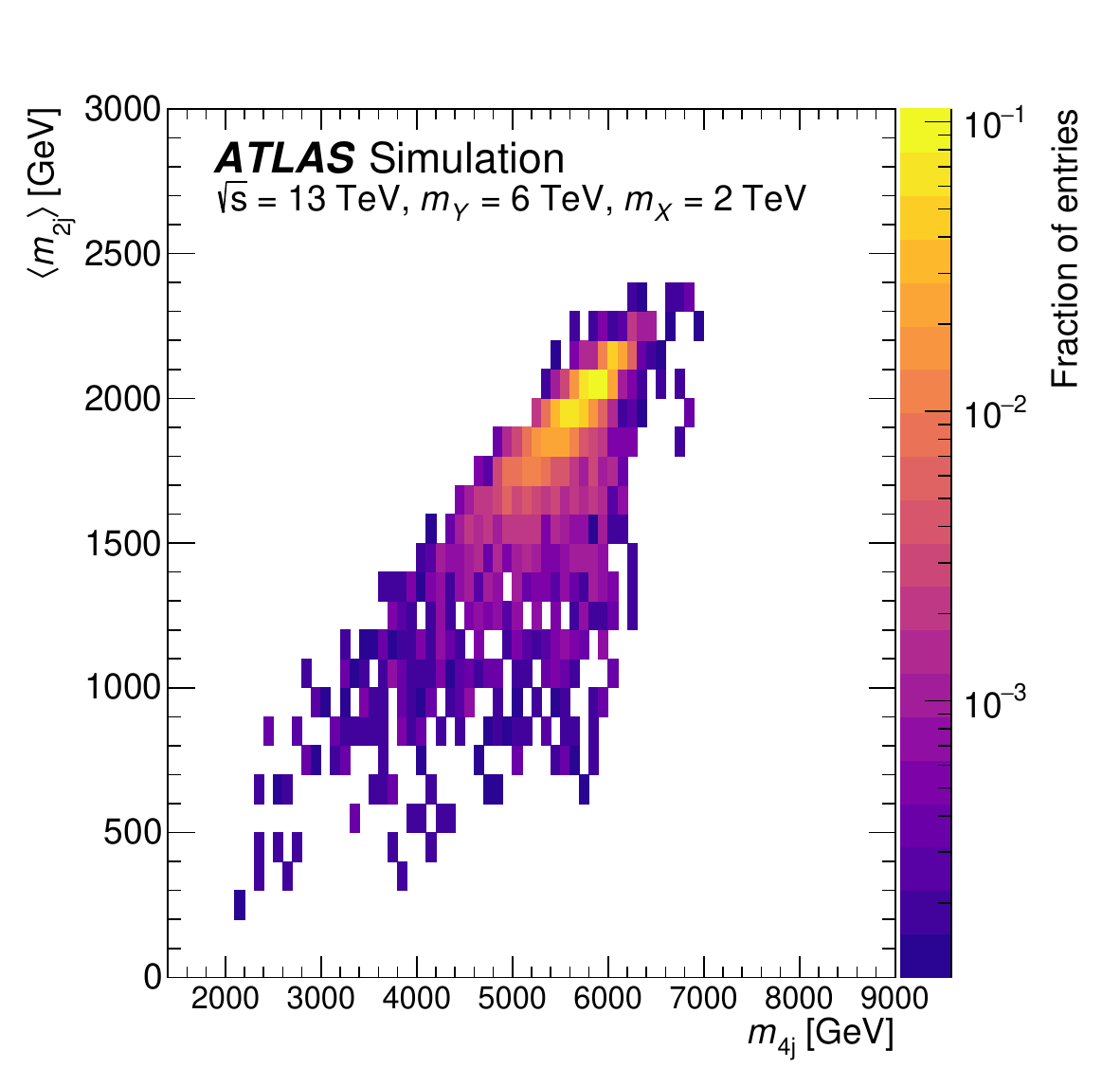}\label{fig:evsel:m4j:m2javg:sig}}\\
\subfigure[]{\includegraphics[width=0.4\textwidth]{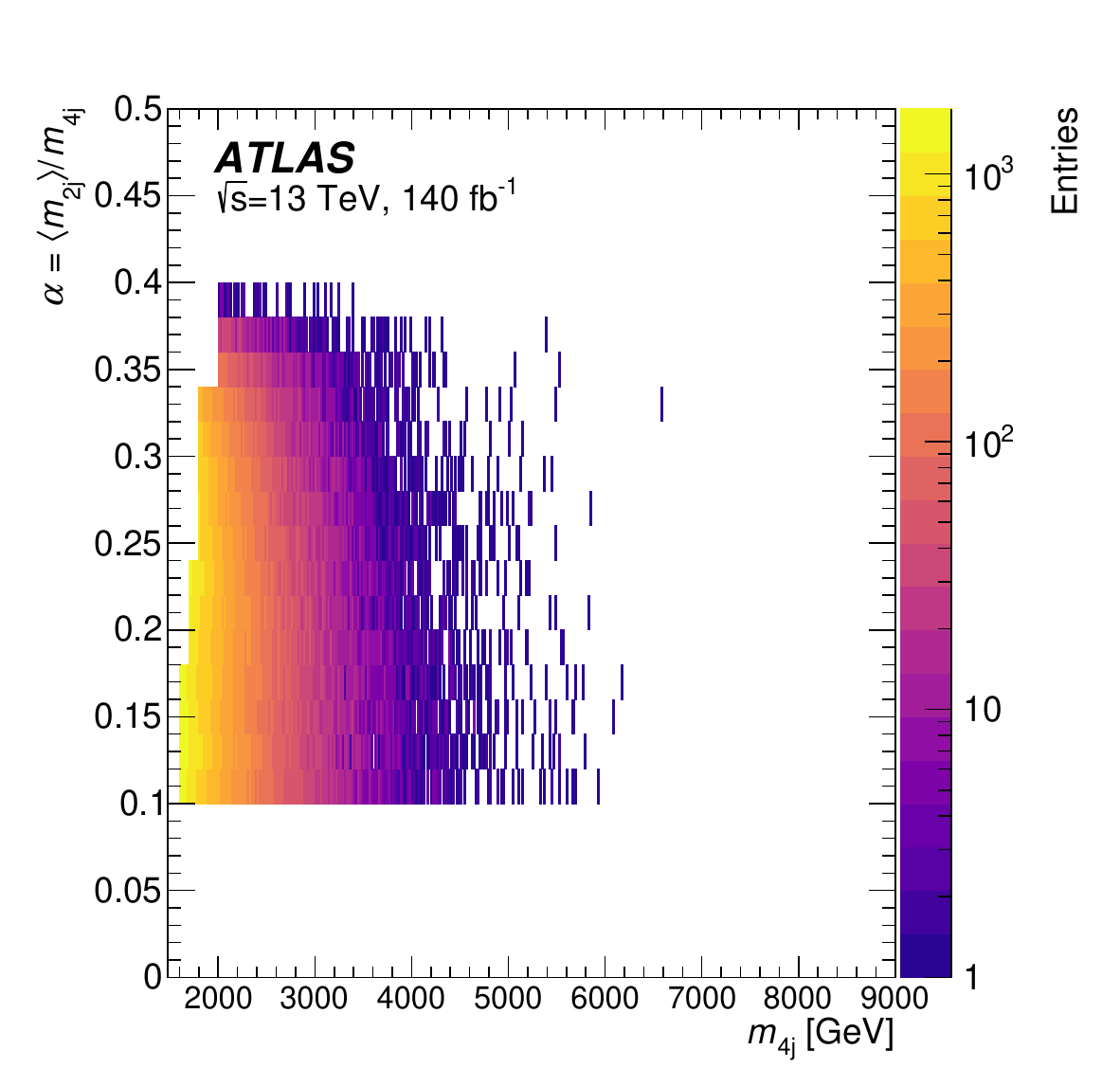}\label{fig:evsel:m4j:alpha:data}}
\subfigure[]{\includegraphics[width=0.4\textwidth]{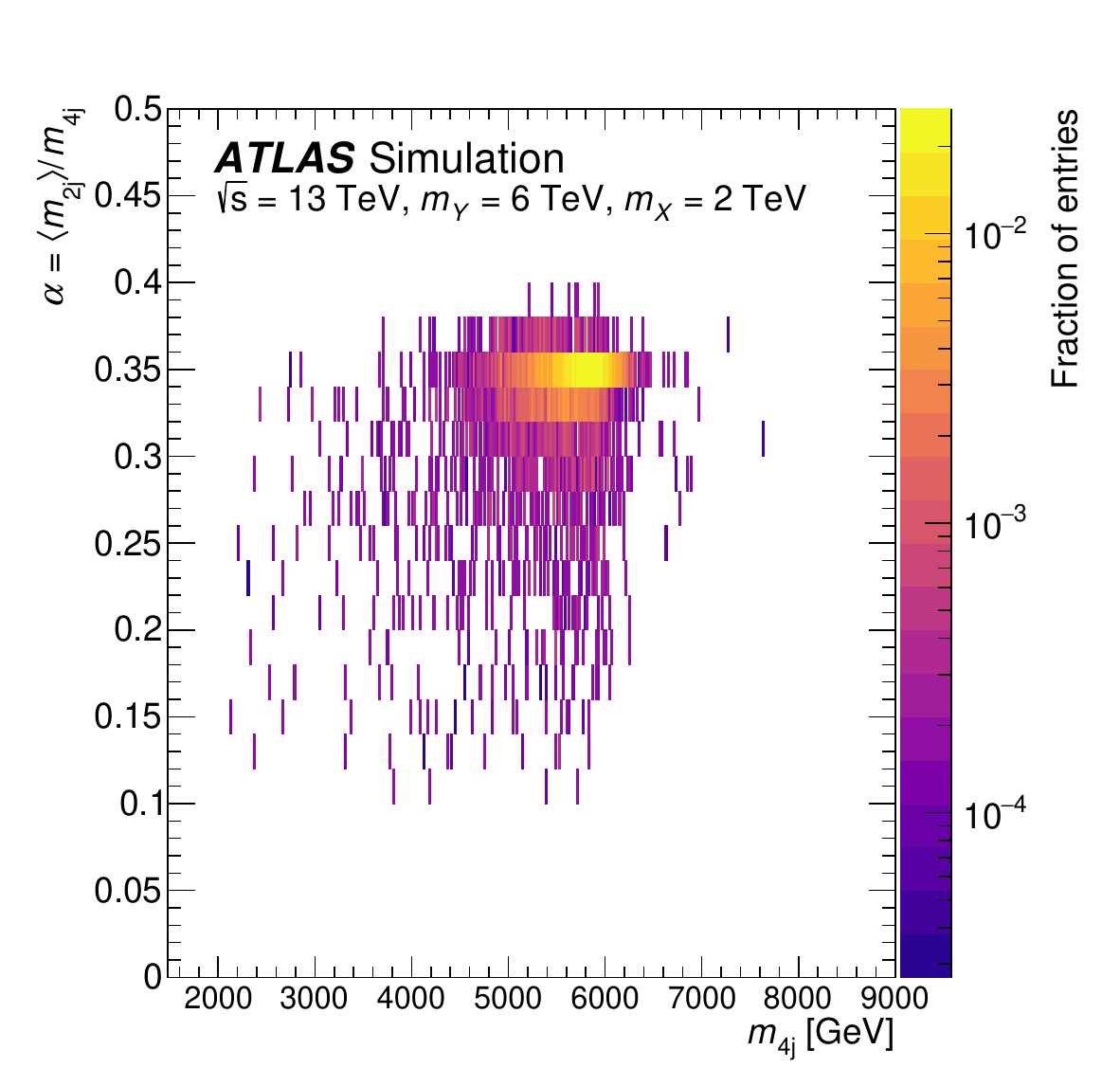}}\\
\subfigure[]{\includegraphics[width=0.4\textwidth]{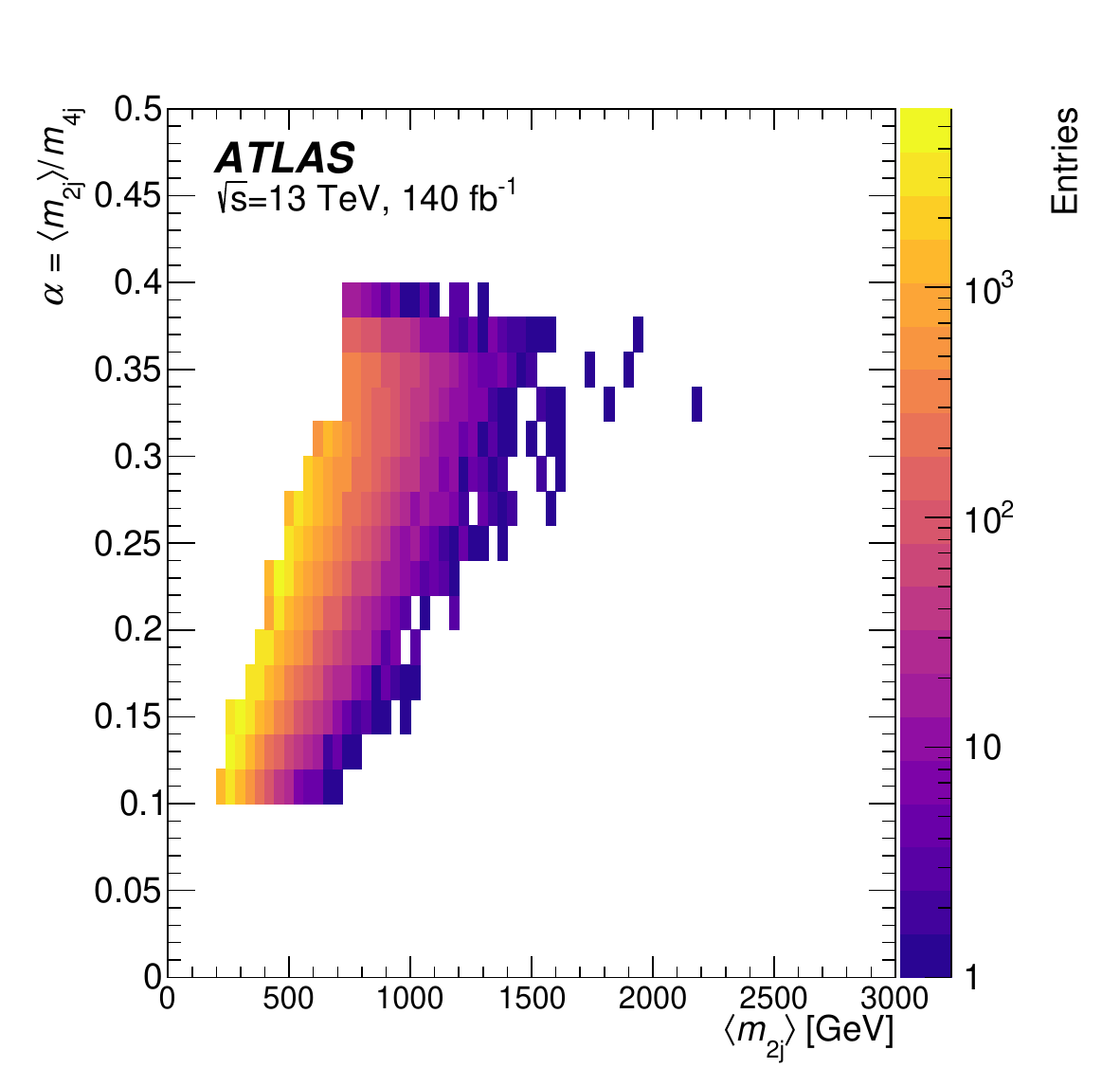}}
\subfigure[]{\includegraphics[width=0.4\textwidth]{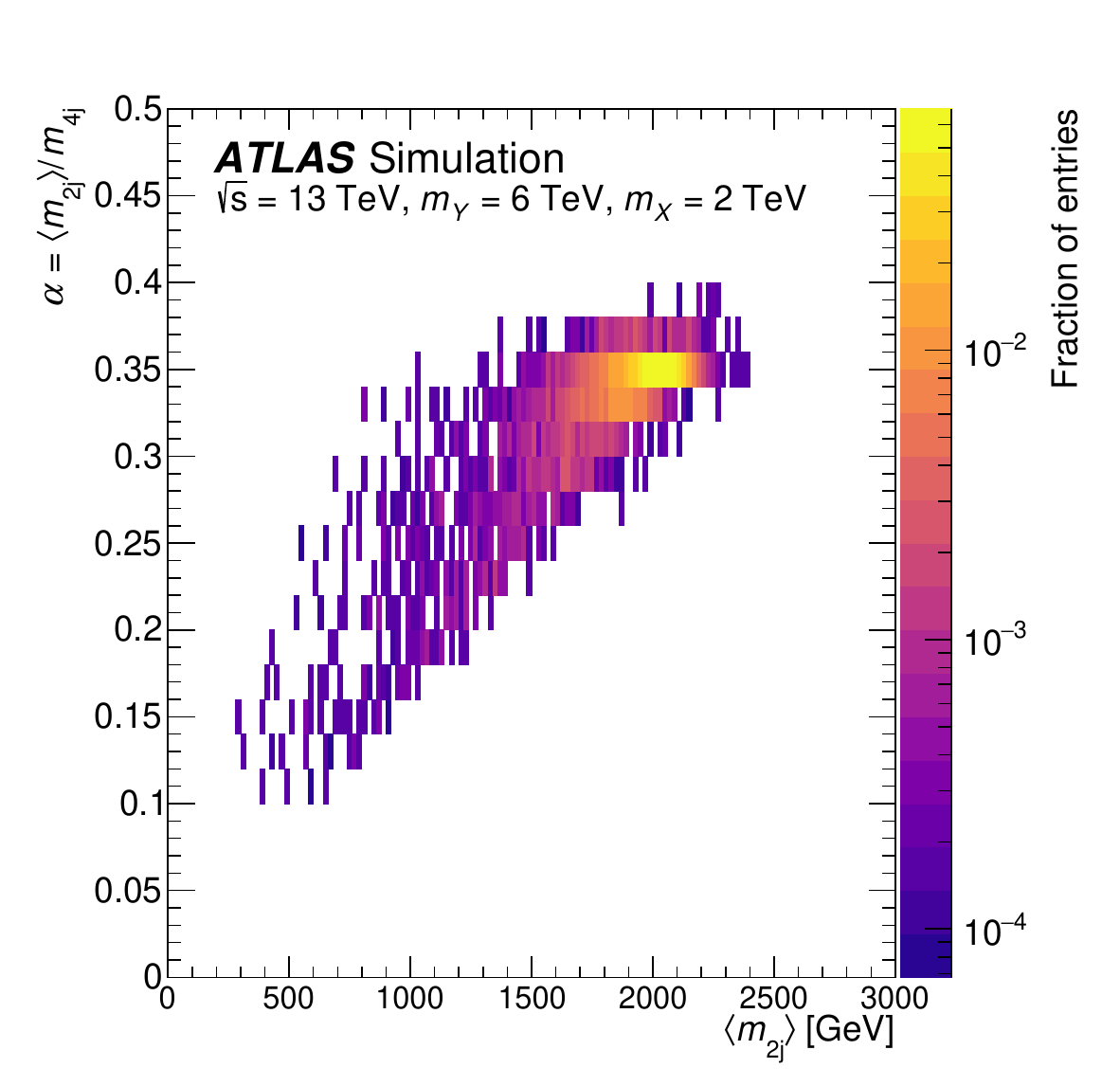}\label{fig:evsel:m2javg:alpha:sig}}
\caption{Two-dimensional histograms of (a, b)~\mtwoavg~\emph{vs.}~\mfour, (c, d)~$\alpha$~\emph{vs.}~\mfour~and (e, f)~$\alpha$~\emph{vs.}~\mtwoavg.
The left column shows the distributions in data, the right column shows the distributions for a simulated signal sample with $m_{Y}=6~\TeV$ and $m_{X}=2~\TeV$.}
\end{figure}
 
Events in data are required to have been selected by one of several single-jet triggers, whose thresholds varied depending on the data-taking period during Run~2.
The particular triggers used to select events in a given run period were always the lowest unprescaled triggers recording data during that time.
The single jet triggers become fully efficient for different values of~\mfour~ and ~\mtwoavg, depending on the $\alpha$ bin that is selected.
To use fully efficient triggers while also retaining sensitivity to the widest range of values, different minimum~\mfour~ and ~\mtwoavg~ thresholds are imposed on the data and simulated signal events used in the interpretation of the search (see Section~\ref{sec:results}).
The trigger thresholds were optimised such that they were at least 99.5\% efficient for each trigger, and are listed in Table~\ref{tab:triggers} for the various  $\alpha$ regions used in the search.
Over 220~000 events in the Run~2 data sample satisfy the analysis selections.
 
\begin{table}
\centering
\caption{Table of selections for the minimum~\mtwoavg~and~\mfour~values considered when selecting events for a given $\alpha$ bin.
These selections are based on a requirement that the single-jet triggers used in the search are at least 99.5\% efficient in the selected region.}\label{tab:triggers}
\begin{tabular}{|c || c | c |}
\hline
$\alpha$ bin & Minimum~\mtwoavg~ & Minimum~\mfour~\\
\hline \hline
$0.10 < \alpha < 0.12$  & 230~\GeV & 1775~\GeV \\
$0.12 < \alpha < 0.14$  & 250~\GeV & 1775~\GeV \\
$0.14 < \alpha < 0.16$  & 270~\GeV & 1725~\GeV \\
$0.16 < \alpha < 0.18$  & 330~\GeV & 1825~\GeV \\
$0.18 < \alpha < 0.20$  & 370~\GeV & 1875~\GeV \\
$0.20 < \alpha < 0.22$  & 430~\GeV & 1875~\GeV \\
$0.22 < \alpha < 0.24$  & 430~\GeV & 1875~\GeV \\
$0.24 < \alpha < 0.26$  & 490~\GeV & 1875~\GeV \\
$0.26 < \alpha < 0.28$  & 510~\GeV & 1925~\GeV \\
$0.28 < \alpha < 0.30$  & 570~\GeV & 1975~\GeV \\
$0.30 < \alpha < 0.32$  & 630~\GeV & 1975~\GeV \\
$0.32 < \alpha < 0.34$  & 730~\GeV & 2175~\GeV \\
\hline
\end{tabular}
\end{table}

\subsection{Signal templates}
 
The results are interpreted using model-independent and model-dependent strategies.
For the model-independent results, the signal templates are modeled as Gaussian distributions, with a mean equal to $m_Y$ and $m_X$ for the ~\mfour~ and ~\mtwoavg~ distributions respectively.
Studying a range of template widths is important, as the theoretical width of a signal can vary across possible signal models.
In this interpretation, template widths ranging from 5\% to 15\% for both~\mfour~and~\mtwoavg were used.
The upper end of the template width is determined from the results of the spurious signal test described in Section~\ref{sec:method:background}.
 
For the model-dependent limits, the shape of the~\mfour~and~\mtwoavg~distributions are parameterized using a Crystal Ball function~\cite{Oreglia:1980cs}, which provides a good description of the shape of mass distribution.
Signal samples are produced with a limited set of signal masses, and these templates are used as inputs to interpolation between mass points to provide a finer signal grid.
The interpolation is done separately for each $\alpha$ region by morphing the parameterized Crystal Ball fits of the signal shape.

\subsection{Background estimation}\label{sec:method:background}
 
Non-resonant QCD processes, which constitute the SM background for this search, result in multijet systems with smoothly falling invariant mass distributions.
To estimate this background in the search regions, a parametric function is fit to the observed data distributions in 1~\GeV bins:
\begin{equation*}
f(x) = p_{1} (1-x)^{p_{2}} x^{p_{3} + p_{4} \ln(x) + p_{5} \ln(x)^2}, \\ x=m/\sqrt{s};
\end{equation*}
where $p_{1}, p_{2}, p_{3}$, $p_{4}$, and $p_{5}$ are the fitted parameters, and $m$ is either~\mfour~or~\mtwoavg.
This function has been succesfully used in a wide variety of resonance dijet and multijet searches by the CDF, CMS, and ATLAS experiments~\cite{CDF:2008ieg,EXOT-2010-01,CMS-EXO-11-015,EXOT-2013-11,EXOT-2015-02,CMS-EXO-16-032,EXOT-2020-15,EXOT-2019-03,CMS-EXO-21-010}.
For the background estimation, a 4-parameter fit is used, where $p_{5}$ is set to zero, while the 5-parameter fit is used to produce pseudodata to validate the fit strategy.
Three-parameter fit functions were also studied but did not have sufficient flexibility to describe the background.
 
The background distribution is fit using a binned, maximum-likelihood fit. In background-only fits, the signal strength is set to zero, while in the signal-plus-background fits, the signal strength is left as a free parameter.
For the model-dependent interpretation, the signal probability density function is defined as the Crystal Ball function fit to the simulated signal events.
For the model-independent interpretations, the signal is parameterized as a Gaussian distribution, where the signal width is set to be a fixed fraction of the signal peak.
 
The data-driven background fitting procedure was validated with MC simulation using several cross-checks, including `spurious signal tests' and `signal injection tests'.
Tests involving signal-plus-background fits are performed using both templates derived from the simulated signal samples described in Section~\ref{sec:signal},
and for model-independent Gaussian signal shapes, using signal widths of  5\%, 10\%, and 15\% of the signal peak.
 
The spurious signal test evaluates whether the fitting procedure is biased in a manner that will produce a non-zero extracted signal when fitting a data sample with no true signal.
This test is performed for a $4$-parameter fit function by performing a signal plus background fit with a Gaussian signal hypothesis of a specified width to 100 pseudodata distributions that are generated from background-only fits to the data distribution with an $5$-parameter function, to provide more flexibility to the background distribution than to the fit function.
Signal widths for both the $X$ and $Y$ ranging from 5\% to 15\% of the signal mean are tested, as well as using the signal templates directly.
For each pseudodata distribution, the number of extracted signal events, $n_{S}$, is determined, and the median value and standard deviation of $n_{S}$ across all pseudodata distributions are taken to be $S_{\mathrm{spur}}$ and $\sigma_\mathrm{spur}$ respectively.
To satisfy the spurious signal requirements, $S_\mathrm{spur} / \sigma_\mathrm{spur}$ is required to be less than 0.5.
For~\mfour~ and ~\mtwoavg, all $\alpha$ regions satisfy this criterion for the signal templates, and for Gaussian signals with widths of 5\%, 10\%, and 15\%.
 
The signal injection test is performed to ensure that the background fit is able to extract a signal component with the expected signal strength.
Simulated signal models with the Gaussian templates with signal widths of 5\% to 15\% and signal templates are included in the fitted background distribution with a given signal cross-section selected to be in the range of 0--5$\sigma$, where $\sigma=n_{S}/\sqrt{n_{B}}$ and the number of signal and background events are determined using a $2\sigma$ window around the injected signal peak in each test.
For these studies, the extracted signal strength is scaled to be the number of signal events within a $2\sigma$ window around the injected signal peak, to match the definition used for the injection.
The injected signals in this study were extracted for 100 pseudodata distributions, with the requirement that the median extracted significance is within $0.5\sigma$ of the injected significance for fits to both the~\mfour~and~\mtwoavg~distributions in all analysis $\alpha$ regions.
All signal templates and Gaussian signals that passed the spurious signal tests also passed the signal injection tests.
 
\subsection{Systematic and statistical uncertainties}\label{sec:method:uncertainties}
 
When interpreting the analysis in terms of candidate signal models, the impact of various experimental and theoretical sources of uncertainty is considered.
The uncertainties in the luminosity and parton distribution functions are included as Gaussian constraints on the yield,
while the uncertainties in the jet energy scale, jet energy resolution, tune, and theoretical renormalization and factorization scale are included as Gaussian constraints on the shape of the distribution.
The uncertainty in the background modeling is implemented as an additional `spurious' signal-like contribution.
 
\paragraph{Luminosity.}
The uncertainty in the combined 2015--2018 integrated luminosity is 0.83\%~\cite{DAPR-2021-01}, obtained using the LUCID-2 detector~\cite{LUCID2} for the primary luminosity measurements.
It is treated as a single normalization uncertainty applied as a scale factor to the signal models.
 
\paragraph{Parton distribution functions.}
The theoretical uncertainty envelope associated with the NNPDF 2.3 LO set of PDFs is propagated through the analysis, where their impact is primarily on the  normalization of the signal events.
The change in analysis selection efficiency is recalculated for each provided PDF variation, and the standard deviation of all such variations is taken as a measure of the systematic uncertainty due to the PDFs.
This uncertainty is a sub-1\% effect for all signal models considered.
 
\paragraph{Jet energy scale and resolution.}
Systematic uncertainties in the $R=0.4$ JES and JER are evaluated using a series of \emph{in situ} measurements and simulation-based techniques, documented in Ref.~\cite{JETM-2018-05}.
Improvements have been made to the component of the jet energy scale uncertainty related to the extrapolation of single-hadron response measurements~\cite{PERF-2015-05,JETM-2018-08} and combined test-beam results~\cite{Adragna:2009zz,ATLASLiquidArgonHEC:2002hwj} into jets.
These uncertainties are reduced by roughly a factor of two compared to those reported in Ref.~\cite{JETM-2018-05}.
Uncertainties due to differences between the gluon-initiated jet energy response of different MC generator setups have also been reduced (`jet flavor response' in Ref.~\cite{JETM-2018-05}), by performing more granular comparisons of the effect of different parton shower and hadronization models on the jet response using the samples documented in Ref.~\cite{ATL-PHYS-PUB-2019-017}.
Following the improved procedure compared with that documented in Ref.~\cite{JETM-2018-05}, the most significant source of uncertainty in the JES now originates from the absolute~\emph{in situ} JES calibration.
 
\paragraph{Variation of initial-state $\alpha_S$ value in the A14 set of tuned parameters.}
The A14 set of tuned parameters used in the \PYTHIA~8 signal simulation includes a pair of `eigentune variations' that can be used to assess the sensitivity of an analysis to the value of the QCD coupling, $\alpha_S$, in initial-state radiation (ISR)~\cite{ATL-PHYS-PUB-2014-021}.
The value of $\alpha_S$ was varied between $0.115-0.140$ from its initial value of $0.127$.
The impact of this variation is negligible compared with other systematic variations.
 
\paragraph{Theoretical renormalization and factorization scale variations.}
The QCD renormalization and factorization scales ($\mu_R$, $\mu_F$) used in the parton shower of the \PYTHIA~8 signal samples are each varied up and down by a factor of two, via weights provided by the \PYTHIA event generator~\cite{Mrenna:2016sih}.
These variations assess the sensitivity of the analysis to parton shower configurations that contain branchings that may compromise the \PYTHIA parton shower's underlying assumptions.
Scale variations for such configurations will result in a large variation for that shower.
The theoretical uncertainties resulting from these scale variations in the mean of the signal distribution are typically less than 0.5\%, and are smaller than the JES uncertainties.
 
\paragraph{Background modeling}
 
A systematic uncertainty to cover potential modeling biases is accounted for using the spurious signal $S_\mathrm{modeling}$.
The value of $S_\mathrm{modeling}$ is determined as the envelope of $|S_\mathrm{spur}|$ over ~\mfour~ and ~\mtwoavg~ respectively.
This is implemented as an additional signal contribution, such that
\begin{equation*}
N_\mathrm{signal}(m_{X,Y}) = \sigma_\mathrm{signal} \mathcal{L} A \epsilon + S_\mathrm{modeling} (m_{X,Y}) \theta_\mathrm{modeling},
\end{equation*}
where $N_\mathrm{signal}(m_{X,Y})$ is the number of extracted signal events at a given $m_{X}$ or $m_{Y}$, $\mathcal{L}$, $A$, and $\epsilon$ are the integrated luminosity, acceptance, and efficiency factors respectively and $\theta_\mathrm{modeling}$ is a nuisance parameter associated with the modeling uncertainty.
The acceptance is defined as the fraction of simulated events at generator level passing the analysis selection cuts, while the efficiency is the fraction of reconstructed events passing the selection.


\FloatBarrier
\section{Results}\label{sec:results}

 
\subsection{Search results}
 
Example tetrajet and average dijet invariant mass distributions in data, together with the corresponding fitted background estimates, are shown in Figure~\ref{fig:bumphunter} for two representative $\alpha$ regions.
The example $\alpha$ regions are selected to show the highest tetrajet invariant mass, and the most significant localized excess observed in data.
The data are well-described by the 4-parameter fit function in all $\alpha$ regions, and the global $\chi^2$ $p$-value ranges from 0.74 to 1.00 for the ~\mfour~spectra, and from 0.08 to 1.00 for the ~\mtwoavg~spectra.
The \textsc{BumpHunter}~\cite{Aaltonen:2008vt,Choudalakis:2011qn} algorithm, as implemented in \textsc{pyBumpHunter}~\cite{Vaslin:2022bds,pyBumpHunter}, is used to quantify the statistical significance of possible resonant signals that may be present in the~\mfour~and~\mtwoavg~distributions.
This is performed using mass bins where the bin width is determined by the mass resolution of ~\mfour~ or ~\mtwoavg~ as a function of the mass, where the mass resolution is determined using a Gaussian fit to the mass response distribution.
The width of the invariant mass window scanned by \textsc{BumpHunter} is varied between two and six resolution bins, and all possible windows of the~\mfour~and~\mtwoavg~distributions are scanned, in each $\alpha$ region.
For each scanned window, \textsc{BumpHunter} evaluates the statistical significance of the observed difference between the data distribution and the background fit.
The \textsc{BumpHunter} $p$-value is defined as the smallest observed probability for the data in a given window to deviate from the background prediction by the observed amount due to a Poissonian fluctuation of the background, using pseudo-experiments generated from the background prediction.
The most significant localized excesses identified by the \textsc{BumpHunter} algorithm are found at 3200~\GeV in the $\alpha$~region from $0.24 < \alpha < 0.26$ for the~\mfour~spectra with a global significance of 0.53 standard deviations, and at 800~\GeV in the $\alpha$~region from $0.26 < \alpha < 0.28$ for the~\mtwoavg~spectra with a global significance of 1.98 standard deviations.
 
\begin{figure}[htbp]
\centering
\subfigure[]{\includegraphics[width=0.48\textwidth]{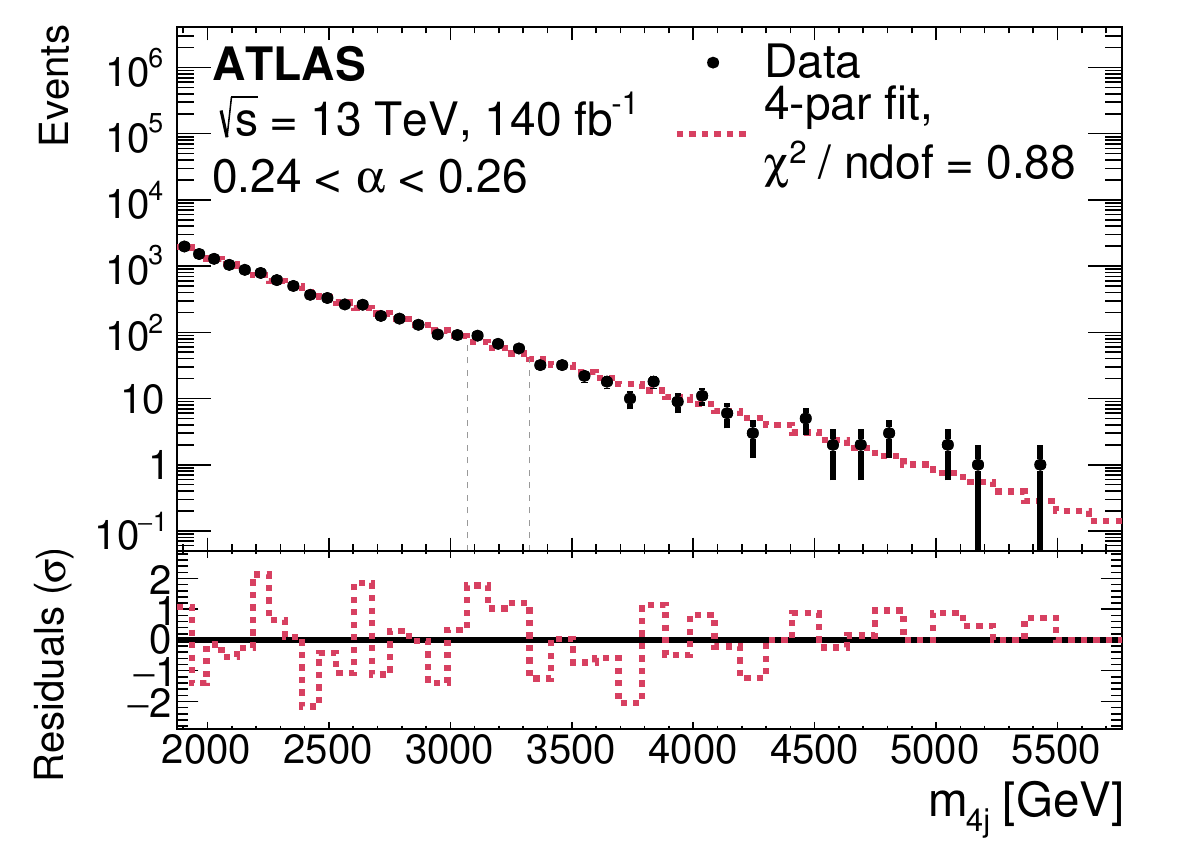}}
\subfigure[]{\includegraphics[width=0.48\textwidth]{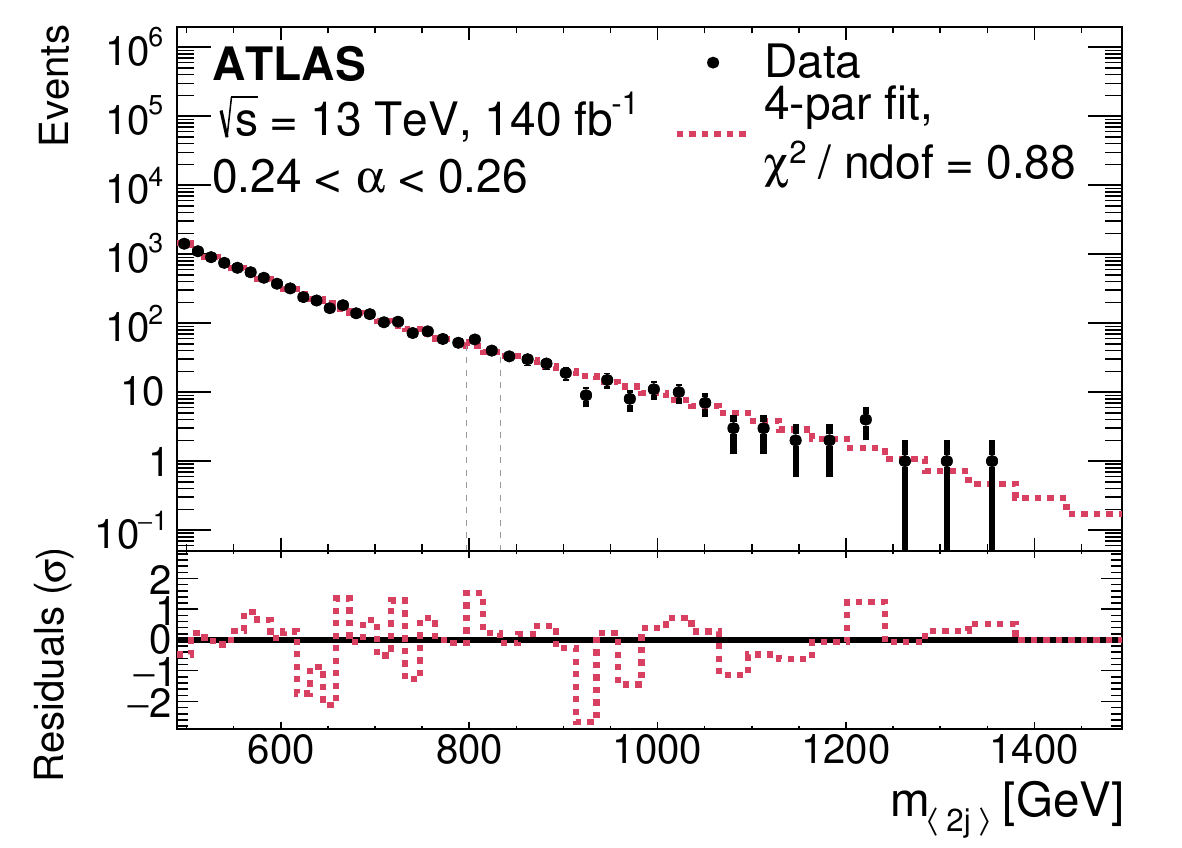}}
\subfigure[]{\includegraphics[width=0.48\textwidth]{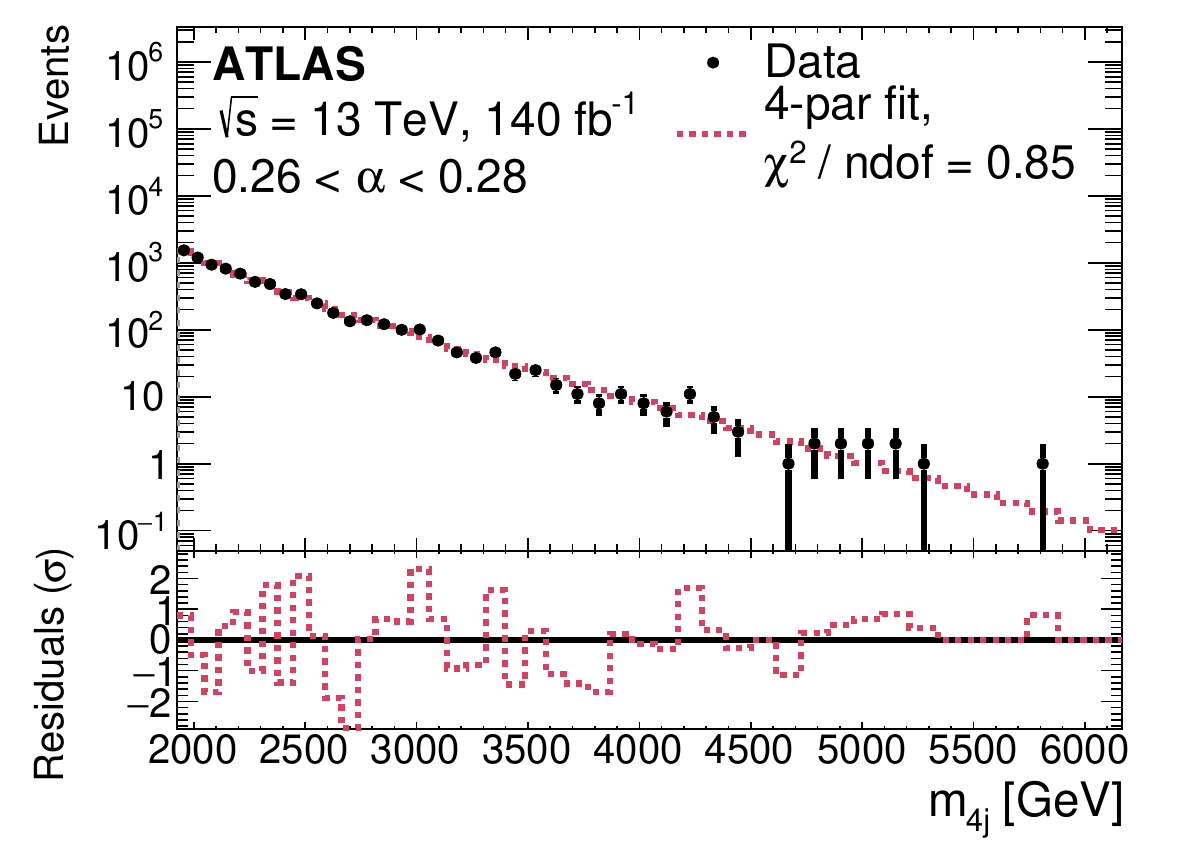}}
\subfigure[]{\includegraphics[width=0.48\textwidth]{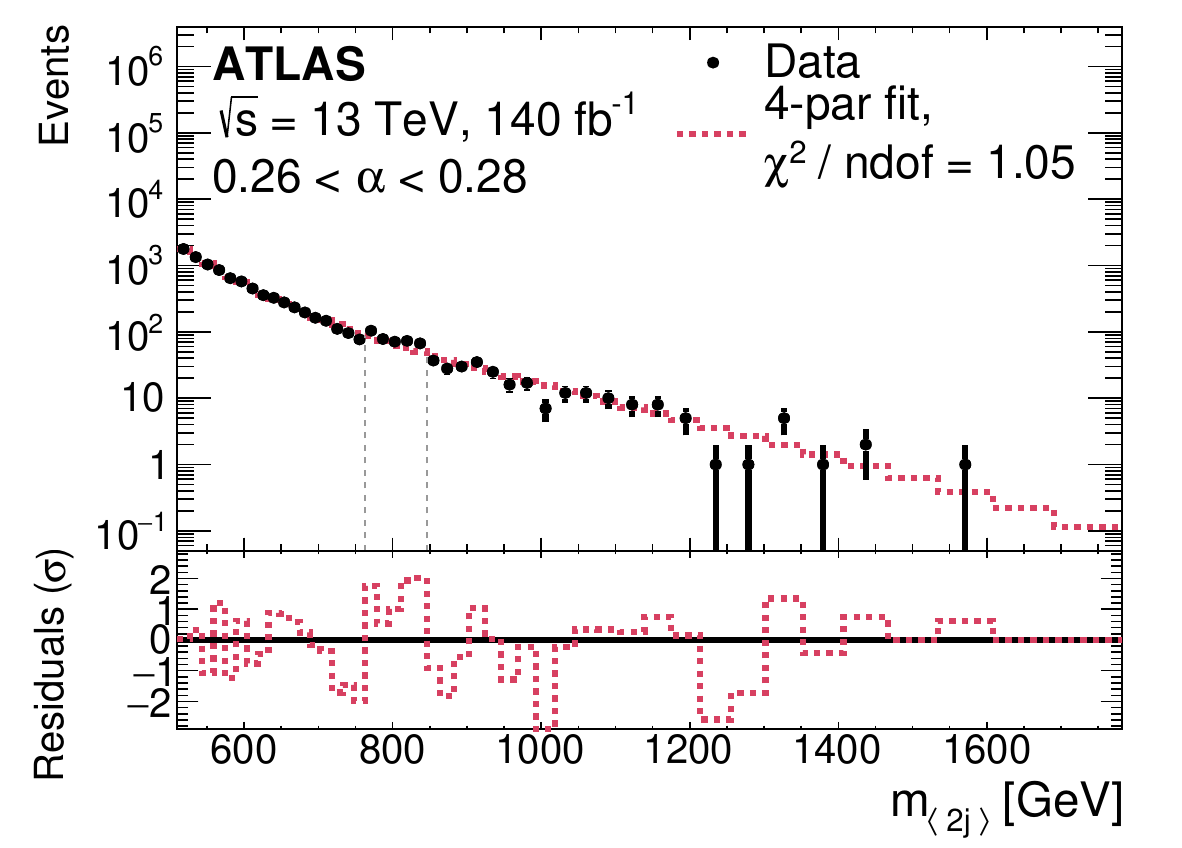}}
\subfigure[]{\includegraphics[width=0.48\textwidth]{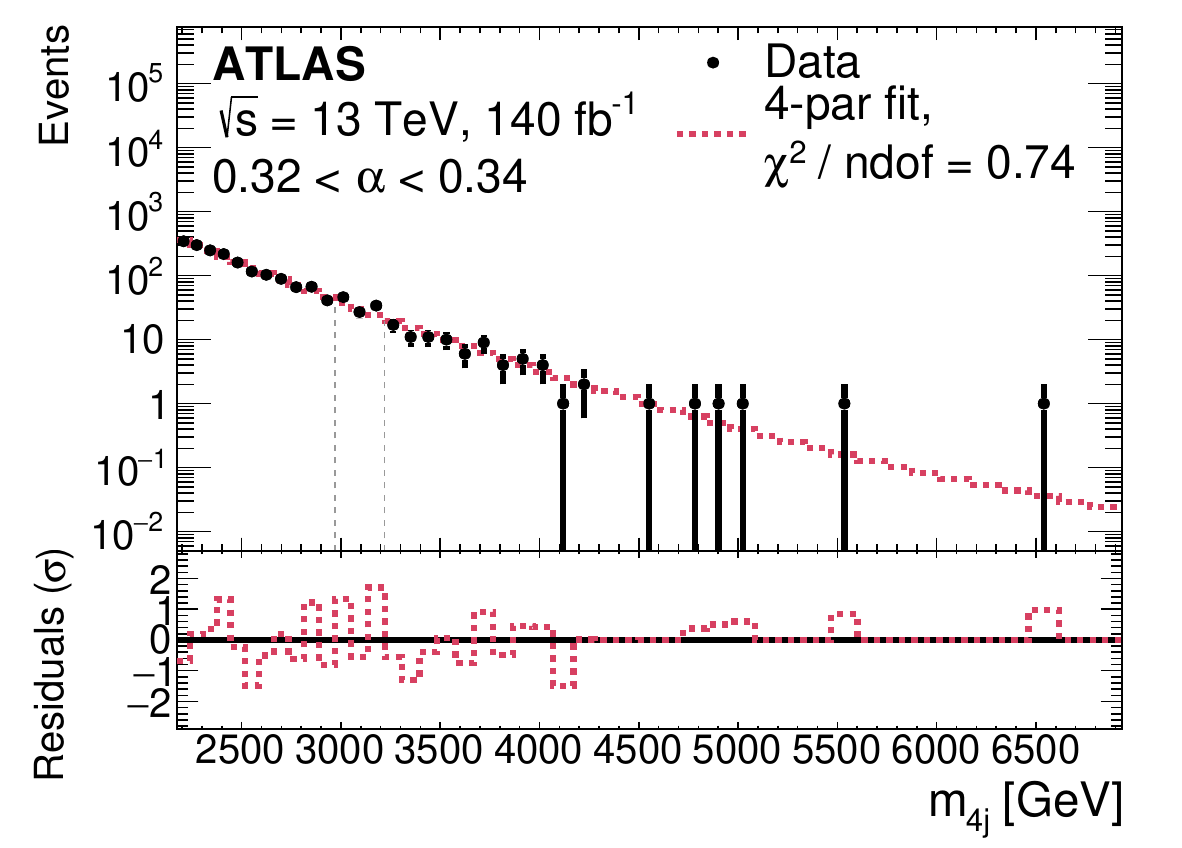}}
\subfigure[]{\includegraphics[width=0.48\textwidth]{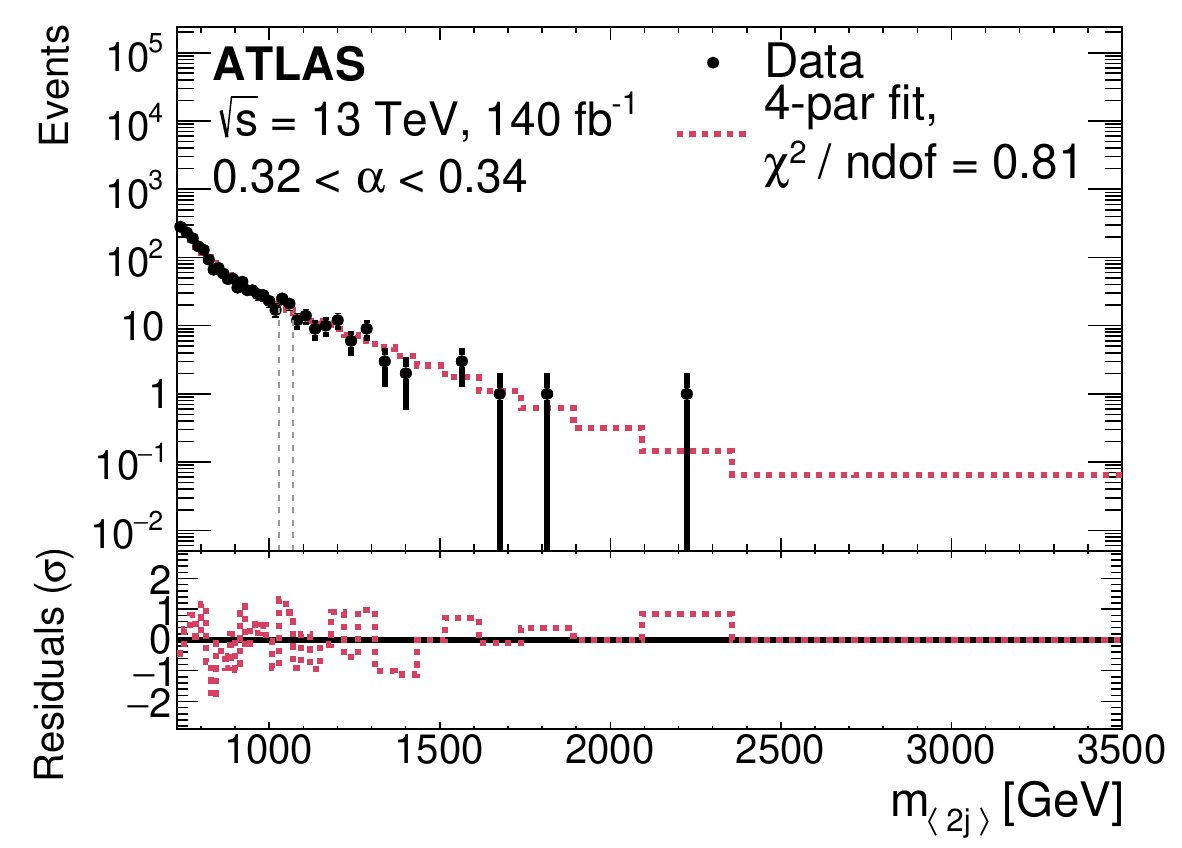}}
\caption{The (a, c, e) tetrajet and (b, d, f) average dijet invariant mass distributions in data are shown, along with the fitted background estimates for (a, b) $0.24 < \alpha < 0.26$, (c, d) $0.26 < \alpha < 0.28$, and (e, f) $0.32 < \alpha  < 0.34$.
The bottom panel of each figure illustrates the fit residuals in terms of standard deviations ($\sigma$).}
\label{fig:bumphunter}
\end{figure}
 
\FloatBarrier
\subsection{Cross-section limits}
 
As no signal is observed, limits can be placed on the range of possible production cross-sections for the hypothetical $Y$ and $X$ bosons.
 
The numbers of signal and background events are estimated from maximum-likelihood fits of the signal-plus-background models to the corresponding~\mfour~and~\mtwoavg~ distributions.
Systematic uncertainties described in Section~\ref{sec:method:uncertainties} are included in the fits via nuisance parameters constrained by Gaussian penalty terms.
The $p$-value is determined from a profile-likelihood-ratio test statistic~\cite{Cowan:2010js}.
The local $p$-value for compatibility with the background-only hypothesis when testing a given signal hypothesis ($p_{0}$) is evaluated based on the asymptotic approximation.
Global significance values are computed from background-only pseudo-experiments to account for the trial factors due to scanning both the signal mass and the width hypotheses.
The expected and observed 95\% confidence level (CL) exclusion limits on the product of the cross-section, branching ratio and acceptance are computed using a modified frequentist-approach CL$_\text{s}$~\cite{Read:2002hq}, in an asymptotic approximation to the test-statistic distribution.
 
Figure~\ref{fig:limits} shows the 95\% CL upper limits on the allowed cross-sections of these particles as a function of their mass, derived using the signal templates used to optimize the analysis for two representative $\alpha$ regions.
Results are interpolated linearly in the logarithm of the cross-section.
Similar results are shown in Figure~\ref{fig:limits:gaus} for the Gaussian signal templates with 5\%, 10\%, and 15\% signal widths.
A summary of the limits for all generated signal masses is shown in Figure~\ref{fig:limits2D} for $m_Y$ and $m_X$ as a function of $\alpha$.
Overall, the limits are smooth as a function of the mass and $\alpha$, and flatten out in the high-mass region where the background estimation predicts significantly less than one event.
 
The relative contribution of statistical and systematic uncertainties (see Section~\ref{sec:method:uncertainties}) on the final analysis sensitivity was assessed by repeating the limit-setting procedure while including only statistical sources of uncertainty.
The analysis sensitivity was not observed to significantly differ during this test.
 
\begin{figure}[htbp]
\centering
\subfigure[]{\includegraphics[width=0.48\textwidth]{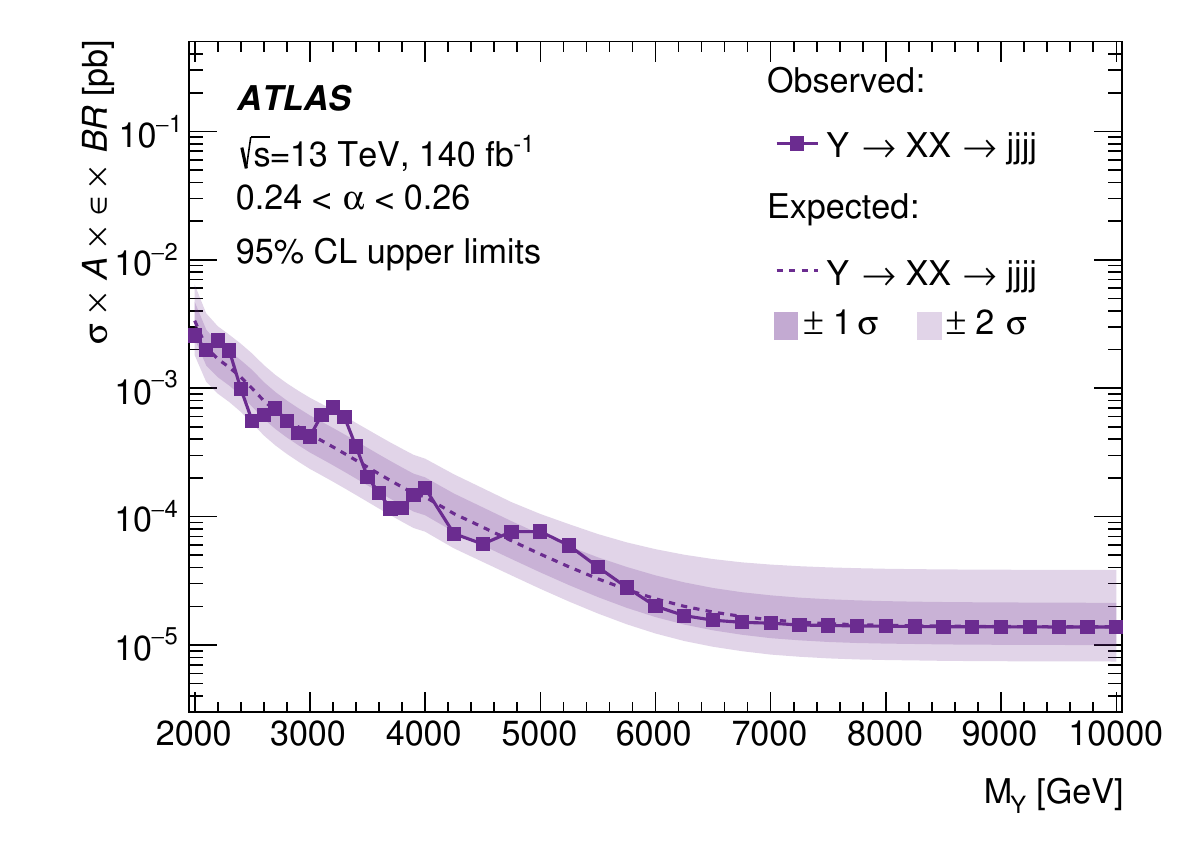}}
\subfigure[]{\includegraphics[width=0.48\textwidth]{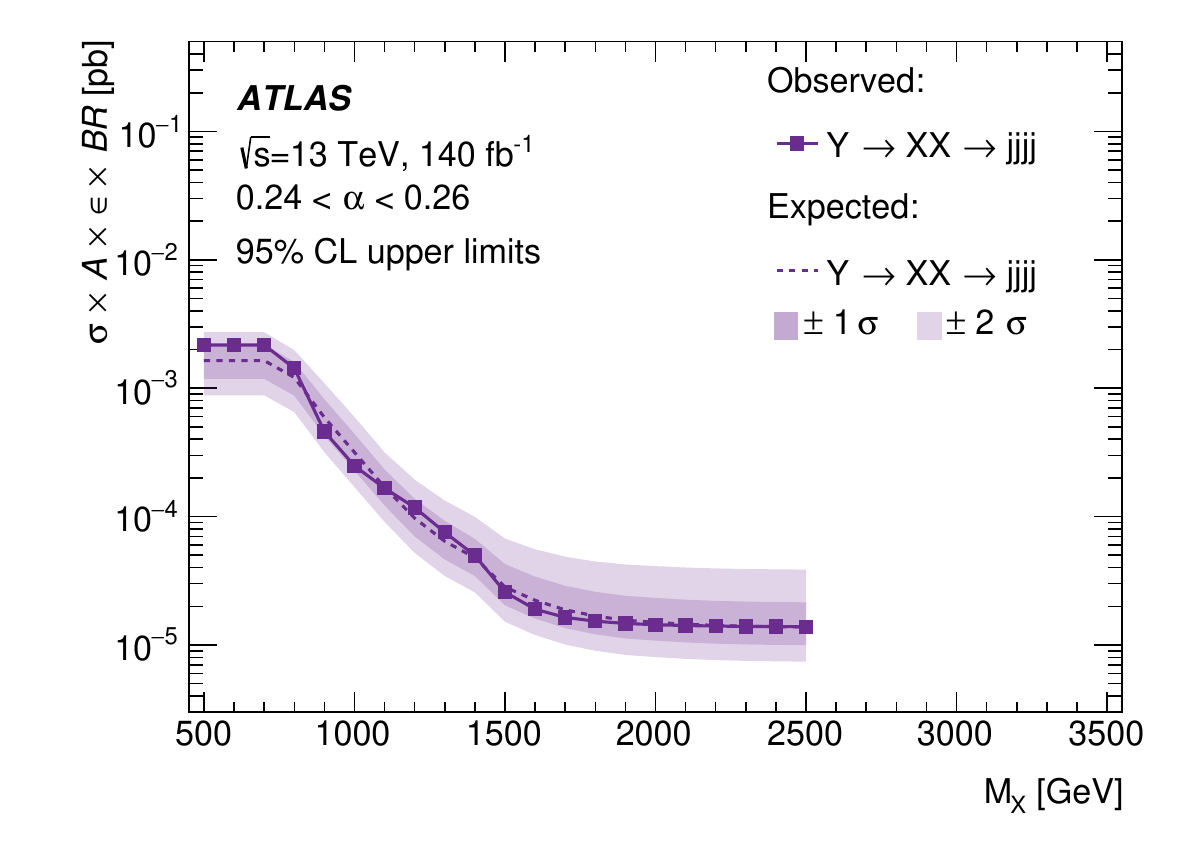}}
\subfigure[]{\includegraphics[width=0.48\textwidth]{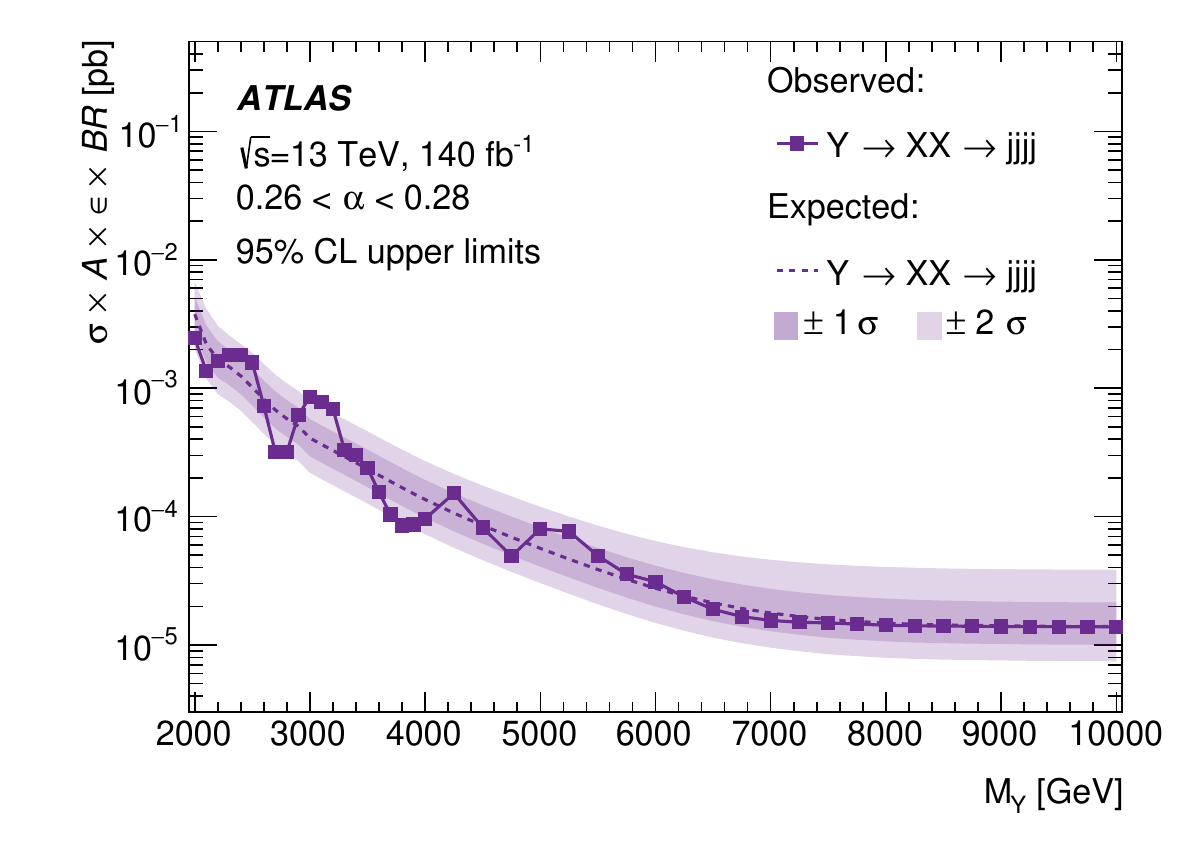}}
\subfigure[]{\includegraphics[width=0.48\textwidth]{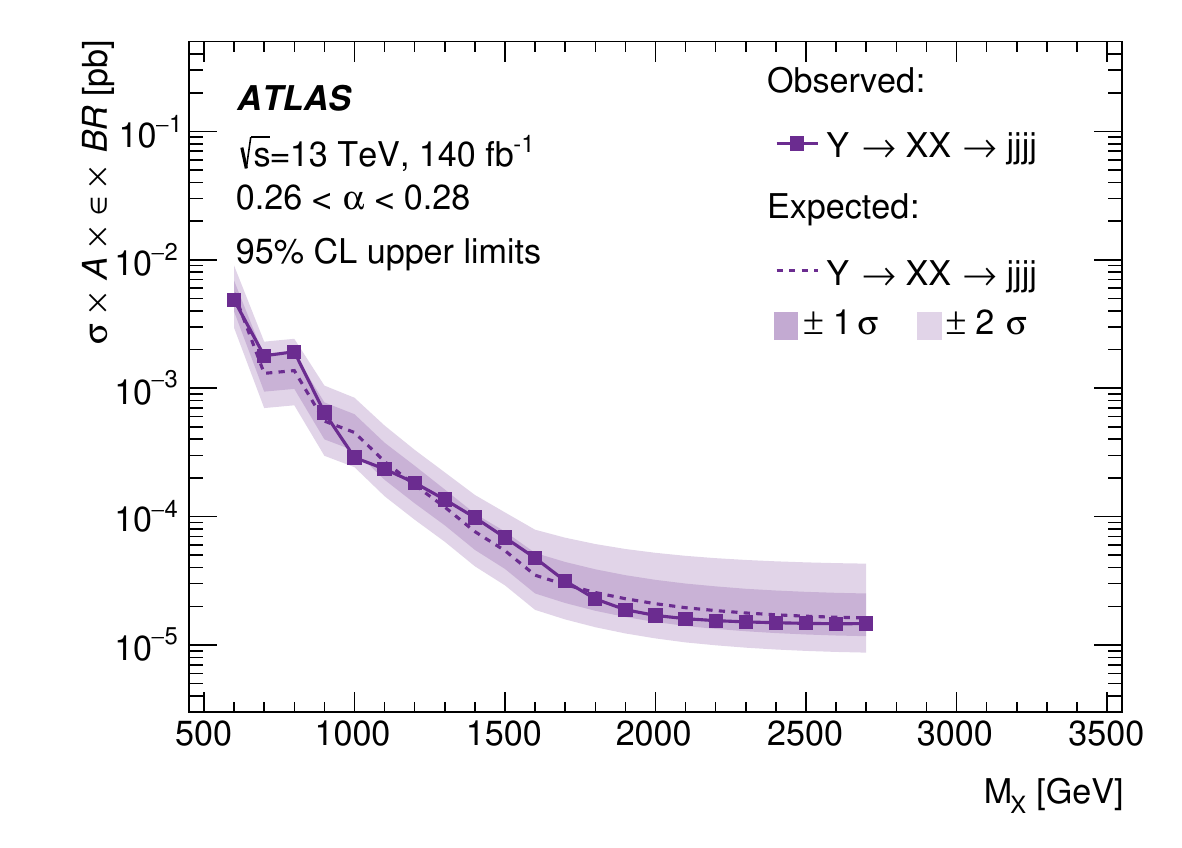}}
\subfigure[]{\includegraphics[width=0.48\textwidth]{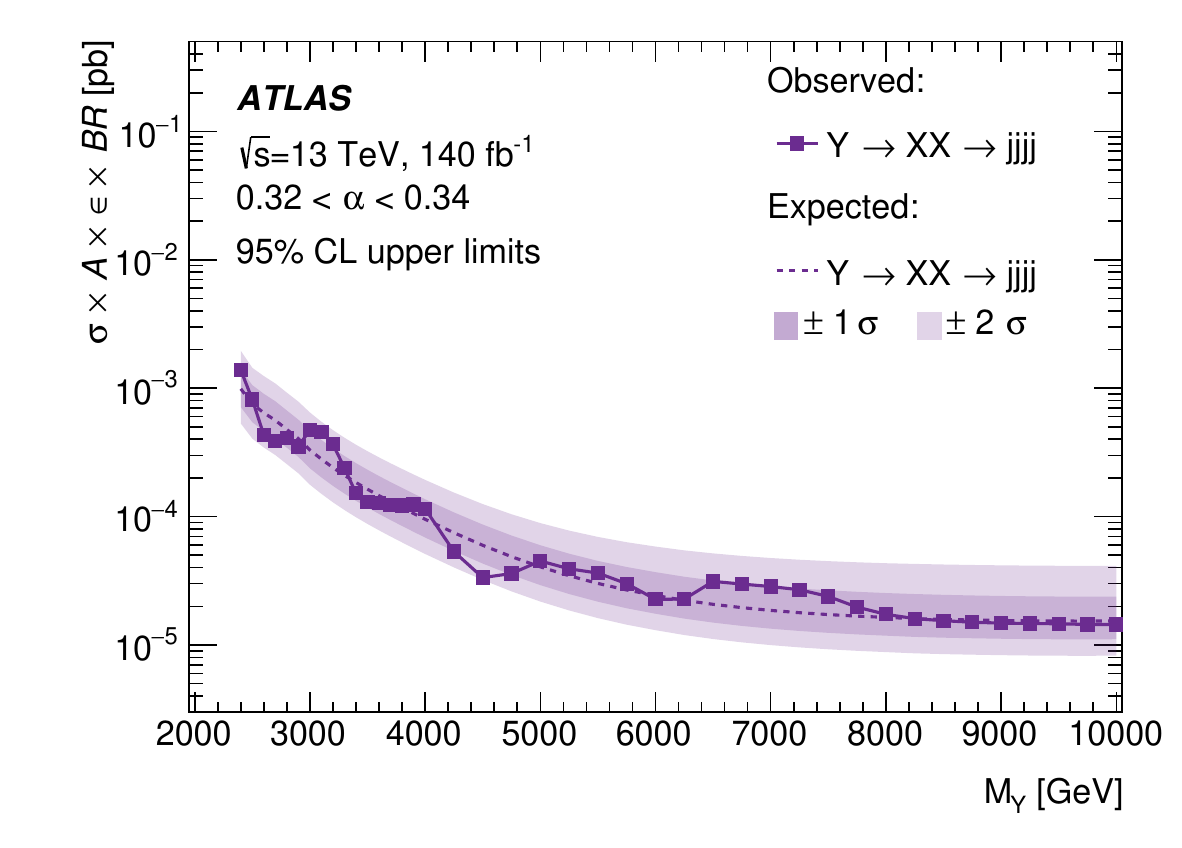}}
\subfigure[]{\includegraphics[width=0.48\textwidth]{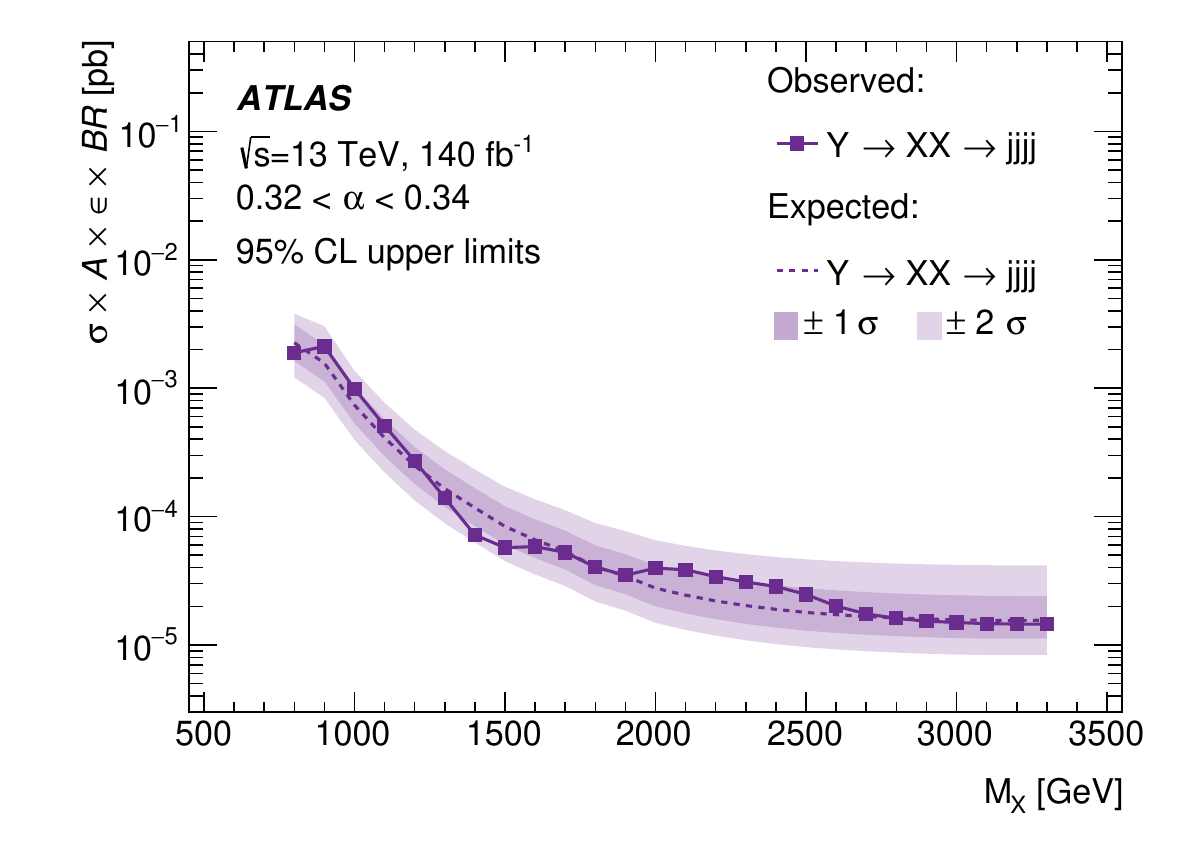}}
\caption{The expected and observed 95\% confidence exclusion limits on the signal cross-section times acceptance ($A$), efficiency ($\varepsilon$), and branching ratio (BR)  as a function of (a, c, e) $m_Y$ and (b, d, f) $m_X$ using the signal templates and a 4-parameter fit function for (a, b) $0.24 < \alpha < 0.26$, (c, d) $0.26 < \alpha < 0.28$, and (e, f) $0.32 < \alpha  < 0.34$.
Observed and expected limits are indicated with markers or a dashed line, respectively.
The shaded bands around the expected limit indicates the (darker band) $1\sigma$ and (lighter band) $2\sigma$ uncertainty range.}
\label{fig:limits}
\end{figure}
 
\begin{figure}[htbp]
\centering
\subfigure[]{\includegraphics[width=0.48\textwidth]{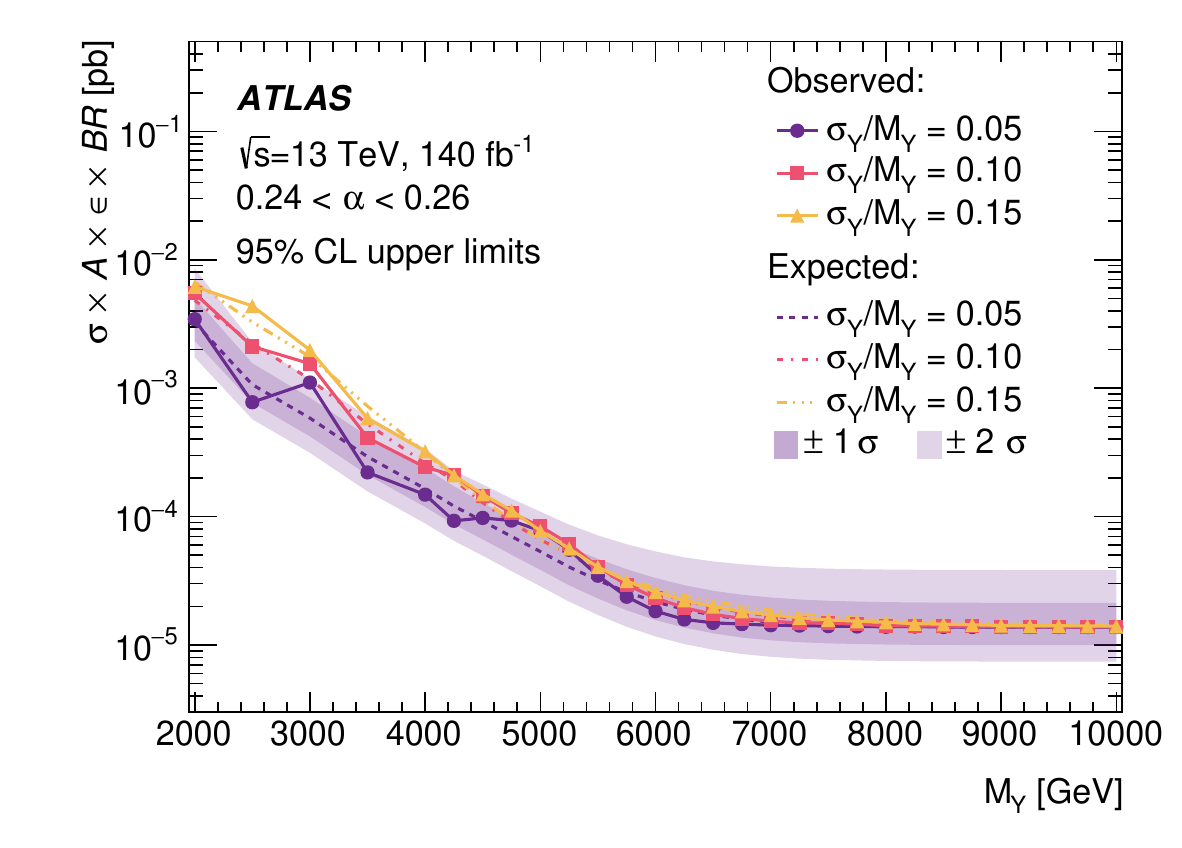}}
\subfigure[]{\includegraphics[width=0.48\textwidth]{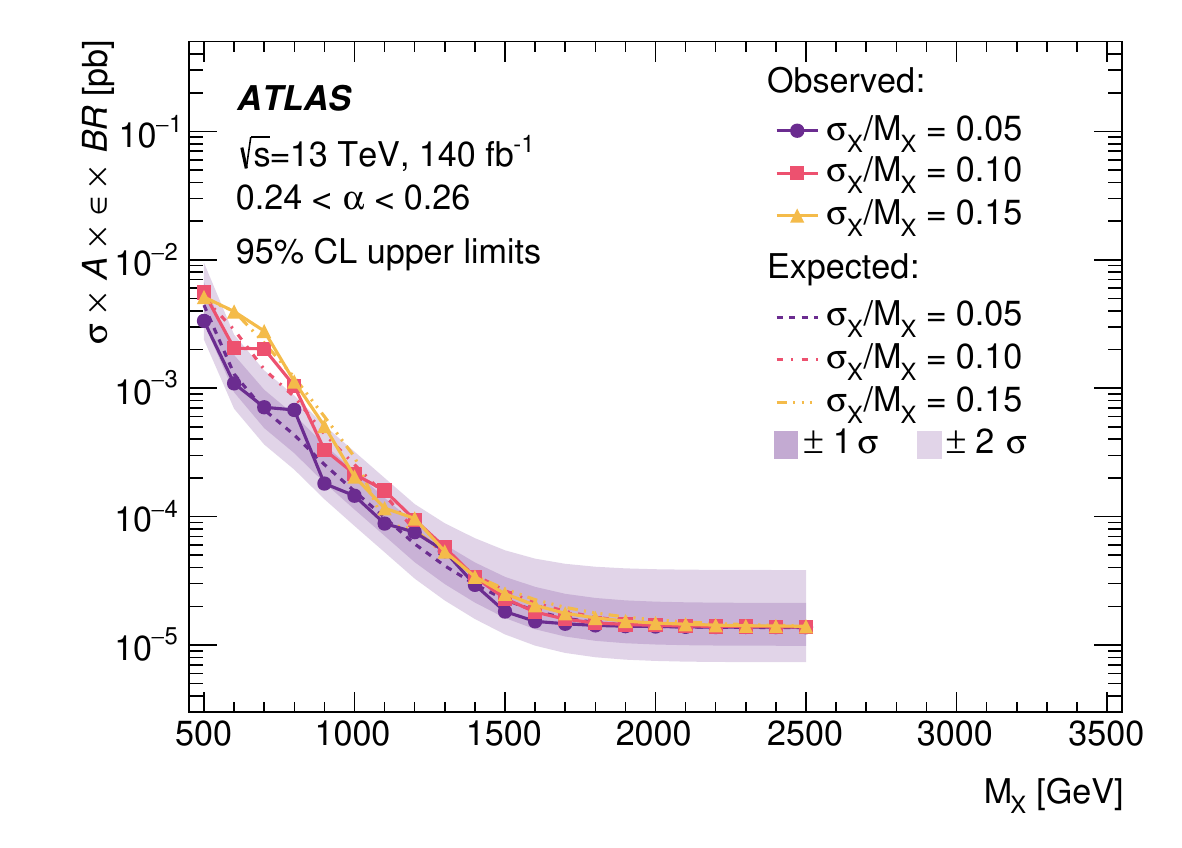}} \\
\subfigure[]{\includegraphics[width=0.48\textwidth]{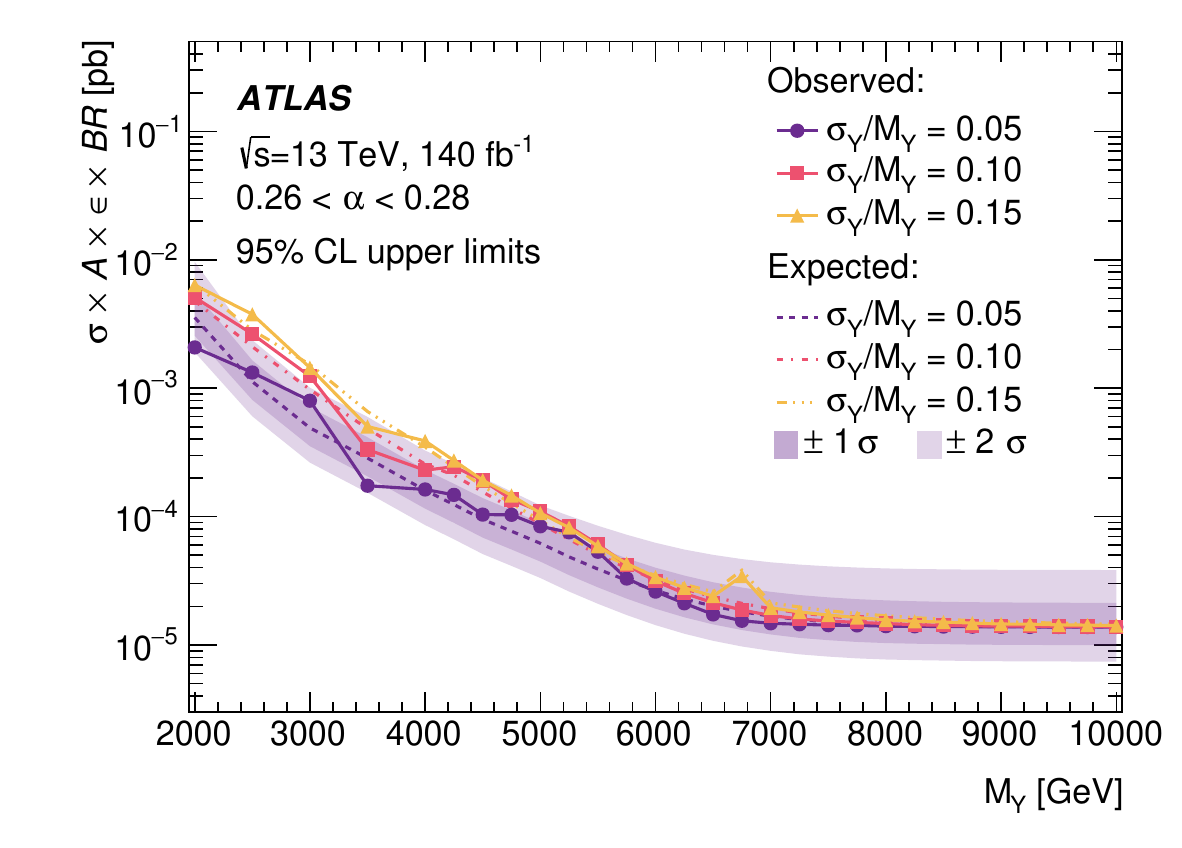}}
\subfigure[]{\includegraphics[width=0.48\textwidth]{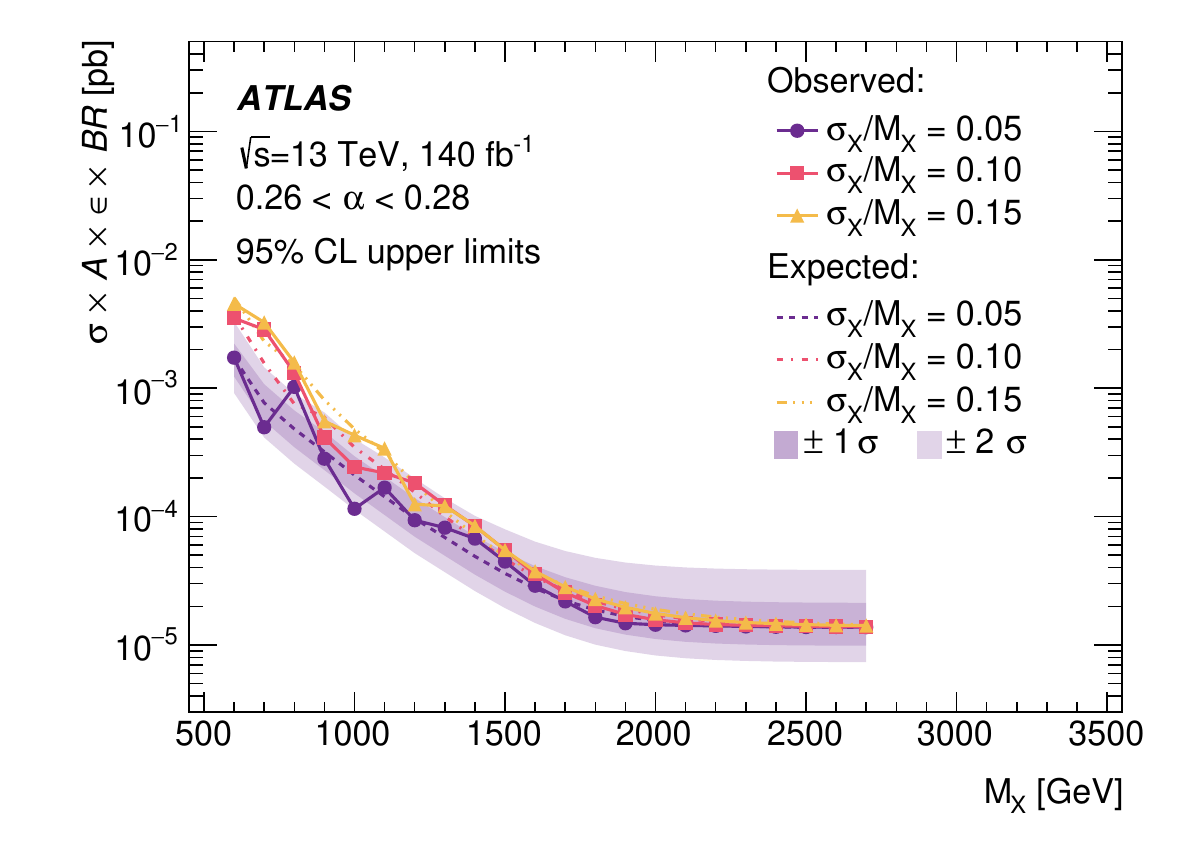}} \\
\subfigure[]{\includegraphics[width=0.48\textwidth]{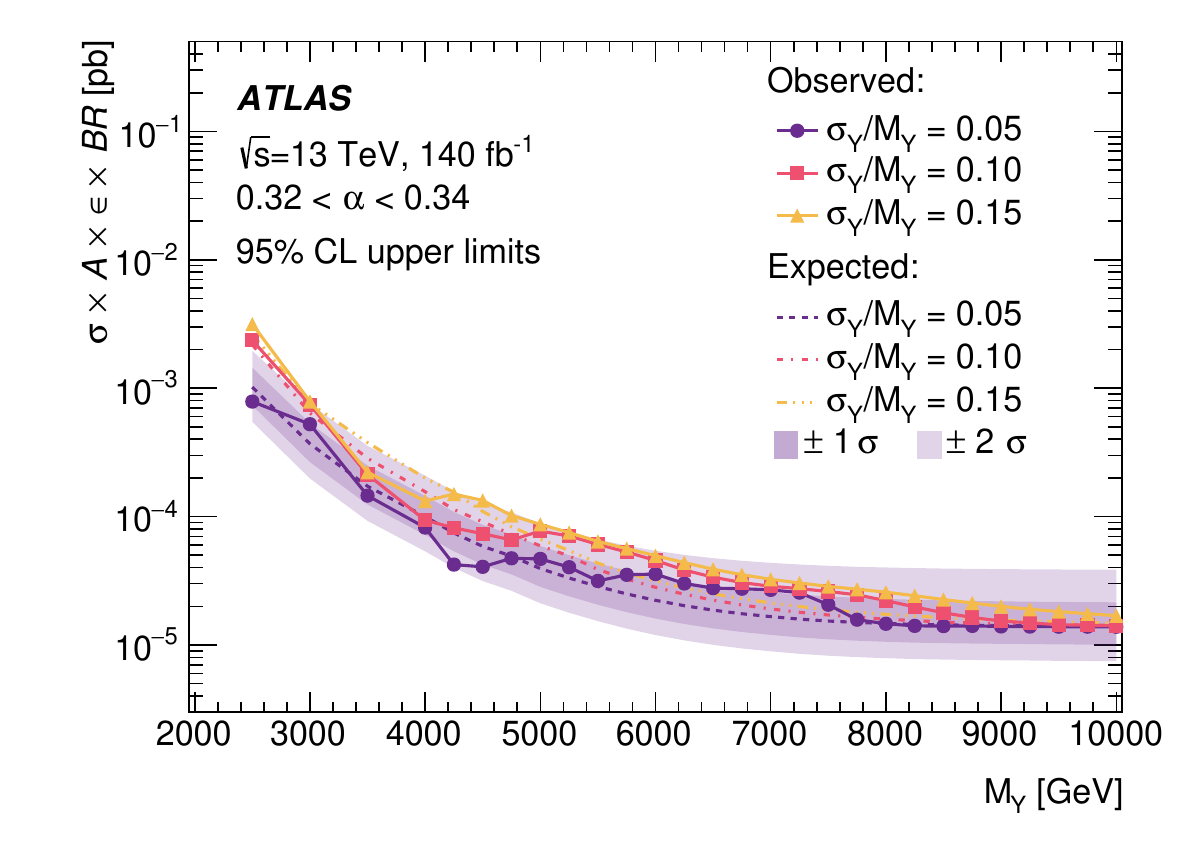}}
\subfigure[]{\includegraphics[width=0.48\textwidth]{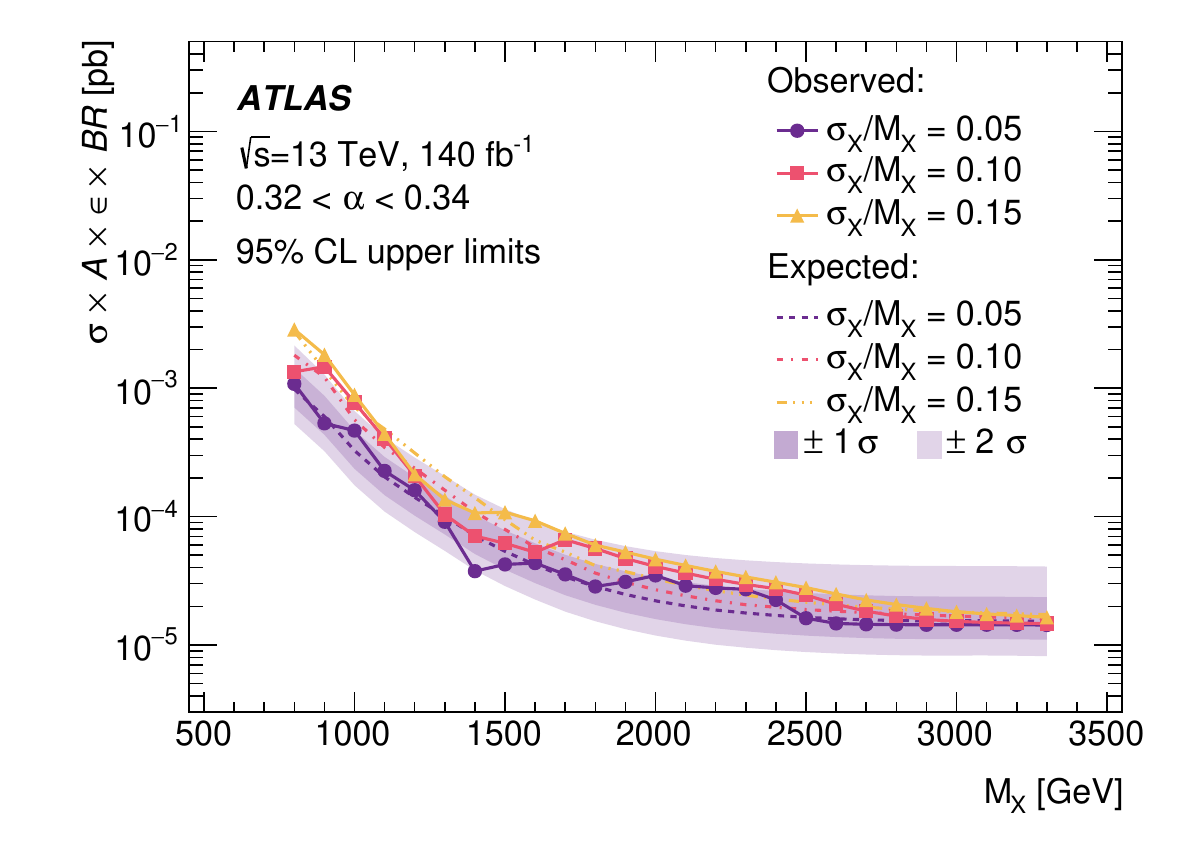}}
\caption{The expected and observed limits on the signal cross-section times acceptance ($A$), efficiency ($\varepsilon$), and branching ratio (BR)  as a function of (a, c, e) $m_Y$ and (b, d, f) $m_X$  for Gaussian signal templates using a 4-parameter fit function for (a, b) $0.24 < \alpha < 0.26$, (c, d) $0.26 < \alpha < 0.28$, and (e, f) $0.32 < \alpha  < 0.34$.
Observed and expected limits corresponding to different choices of template widths (5\%, 10\% and 15\%) are indicated as different sets of markers or line styles, respectively.
The shaded bands around the expected limit for templates with 5\% width indicate the (darker band) $1\sigma$ and (lighter band) $2\sigma$ uncertainty range.}
\label{fig:limits:gaus}
\end{figure}
 
\begin{figure}[htbp]
\centering
\subfigure[]{\includegraphics[width=0.75\textwidth]{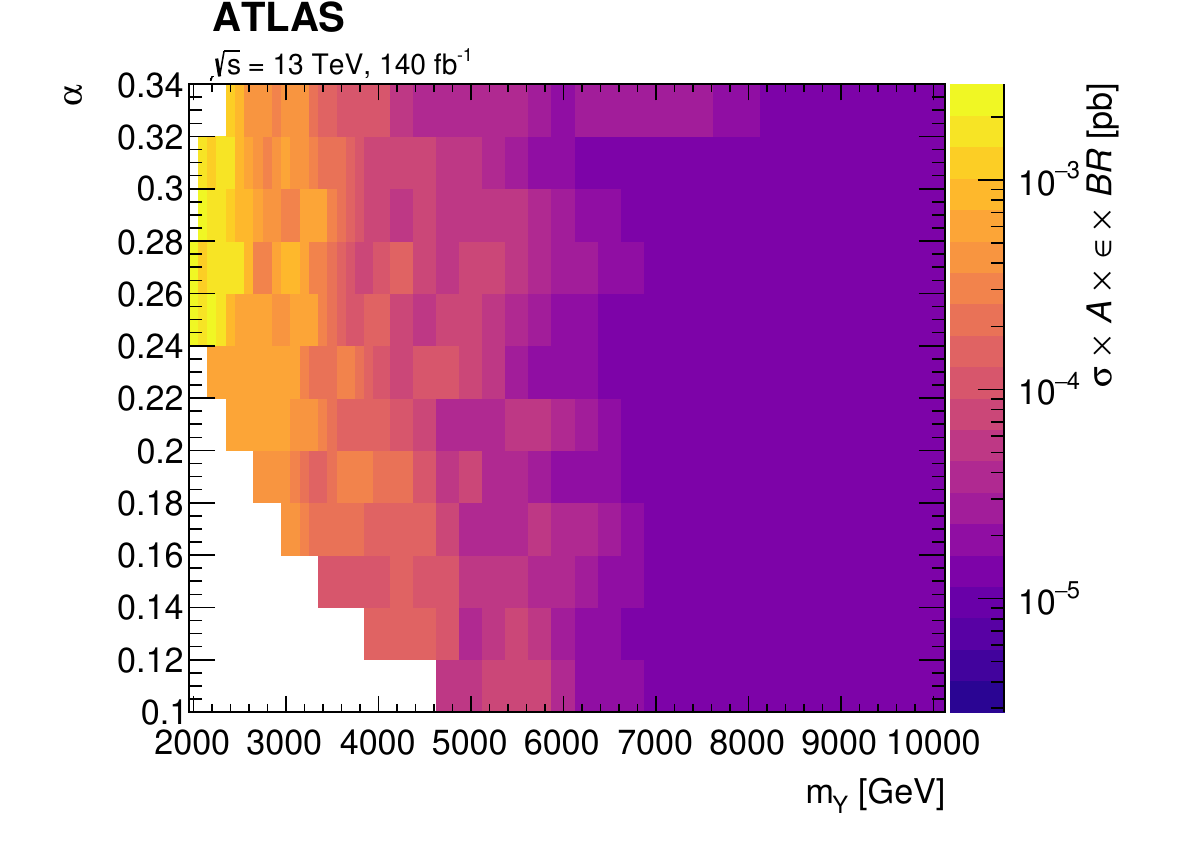}}\\
\subfigure[]{\includegraphics[width=0.75\textwidth]{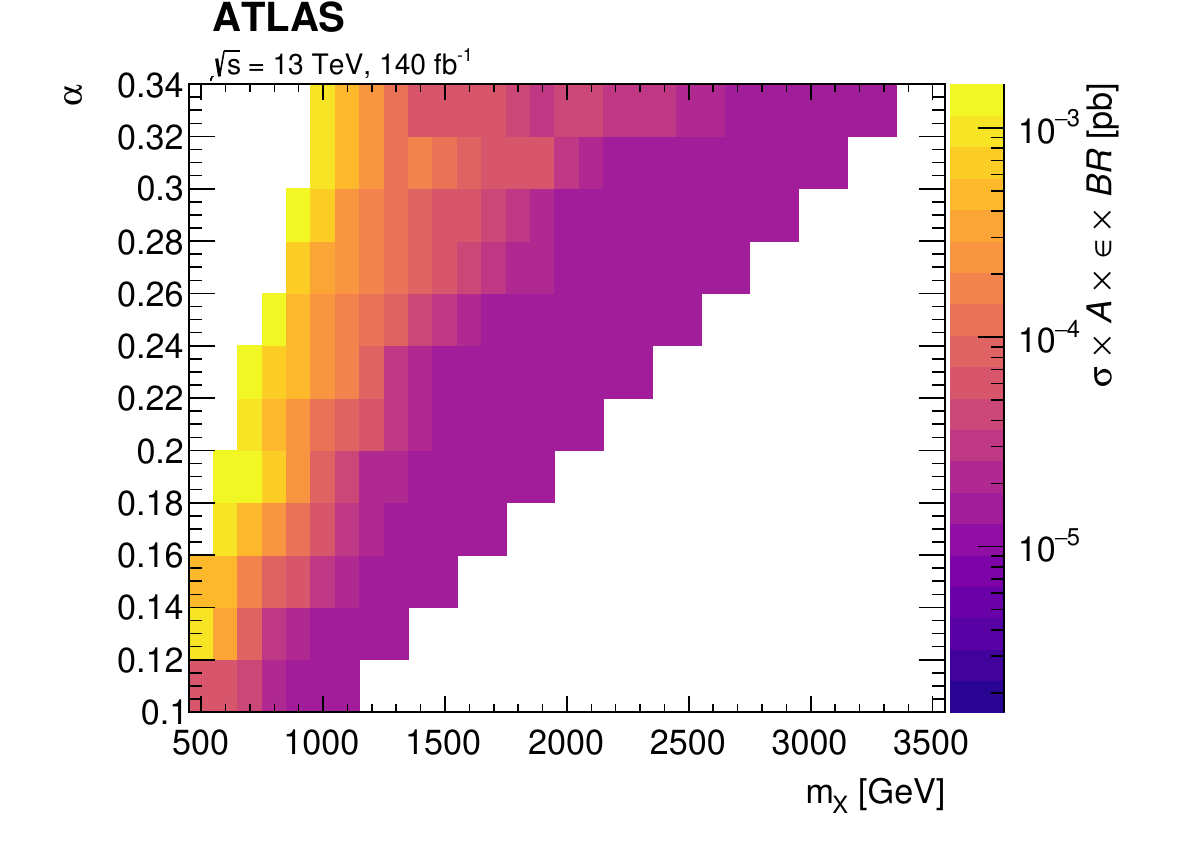}}
\caption{The observed 95\% confidence exclusion limits on the signal cross-section times acceptance ($A$), efficiency ($\varepsilon$), and branching ratio (BR)  as a function of (a) $m_{Y}$, and (b) $m_{X}$ for signals templates using a 4-parameter fit function.}
\label{fig:limits2D}
\end{figure}
 
To better illustrate different types of events passing the event selection, two event displays are shown in Figure~\ref{fig:results:displays}.
The first shows the event with the highest four-jet mass, with a value of $\mfour=6.6~\TeV$ and corresponding~$\mtwoavg=2.2~\TeV$, while the second shows the event with the highest-\pt fourth jet that is selected ($\mfour=5.2$~\TeV,~$\mtwoavg=0.90$~\TeV).
 
\begin{figure}[htbp]
\centering
\subfigure[]{\includegraphics[width=0.70\textwidth]{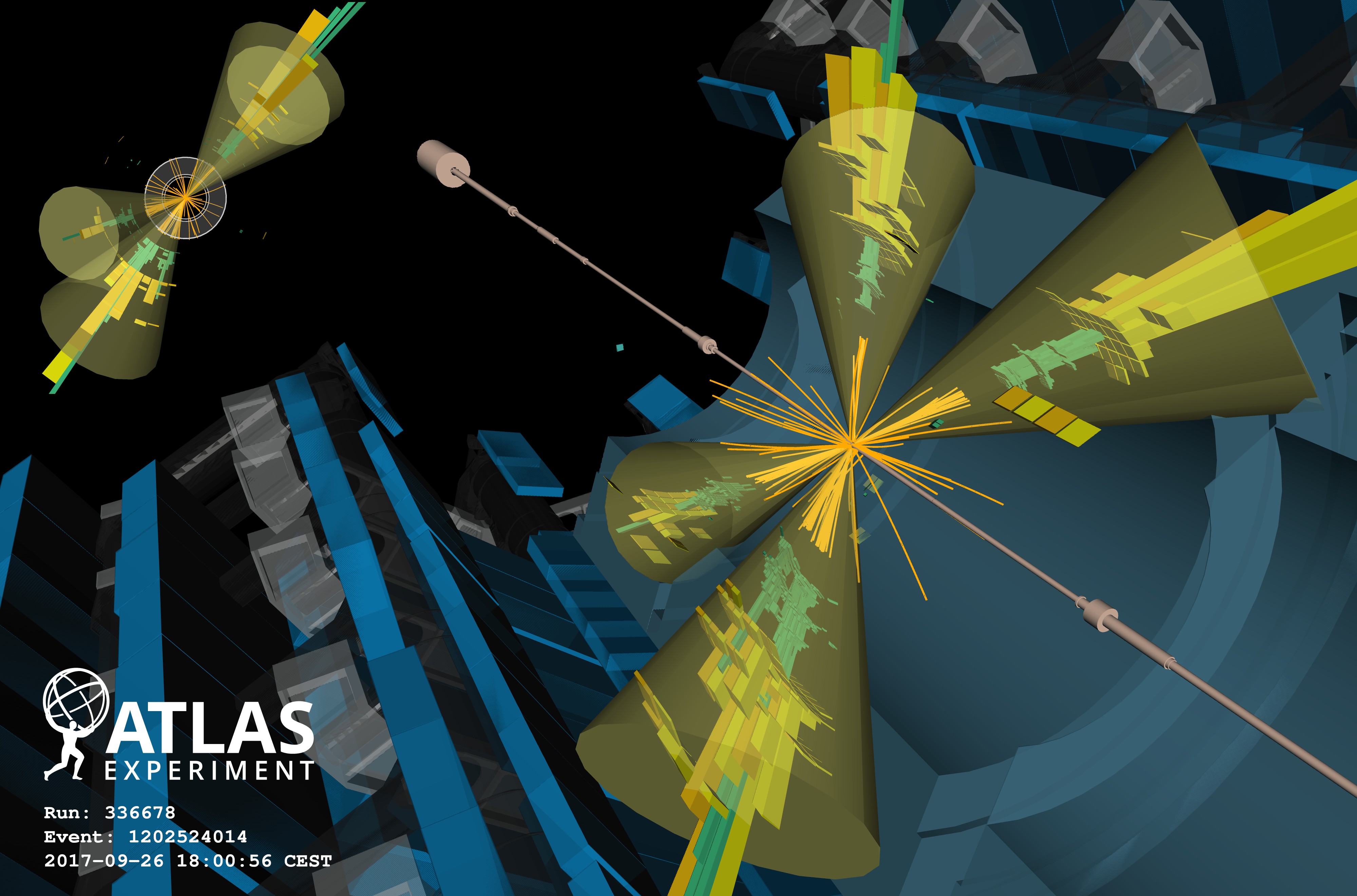}}\\
\subfigure[]{\includegraphics[width=0.70\textwidth]{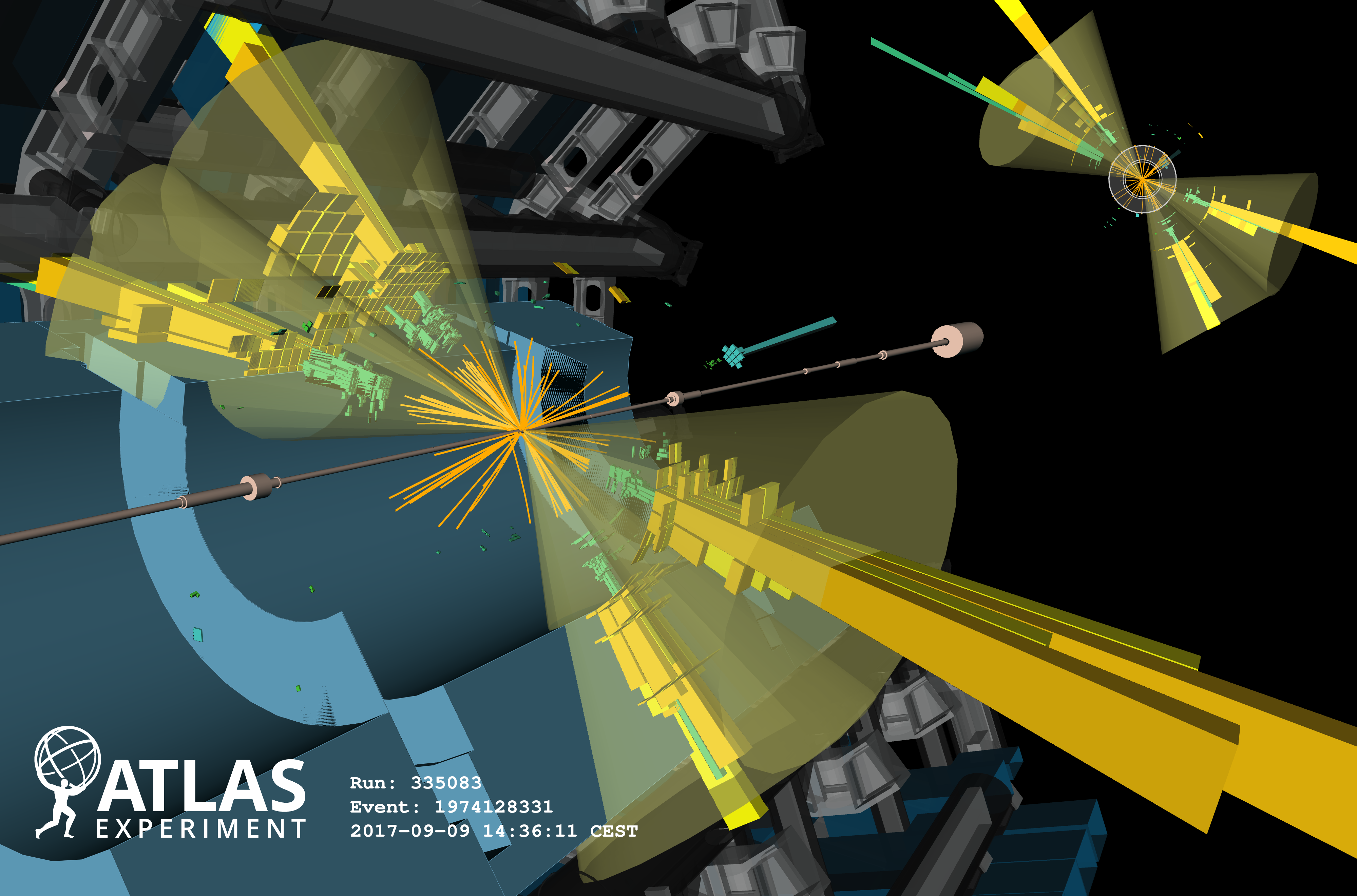}}
\caption{Event display of multijet events (a) Run 336678, Event 1202524014 and (b) Run 335083, Event 1924128331 from proton–proton collisions recorded by ATLAS with LHC stable beams at a collision energy of 13~\TeV.
The former event is displayed from a side-view where the beamline runs horizonally across the image, while the latter event is displayed in a transverse view, down the beamline.
Event (a) possesses the largest tetrajet invariant mass observed in the search ($\mfour=6.6$~\TeV,~$\mtwoavg=2.2$~\TeV), and the \pt values of the four selected jets are 2~\TeV, 1.2~\TeV, 1.2~\TeV~and 0.5~\TeV.
Event (b) is the event with the highest-\pt fourth-jet passing the event selection ($\mfour=5.2$~\TeV,~$\mtwoavg=0.90$~\TeV), and the \pt values of the four selected jets are 1.6~\TeV, 1.3~\TeV, 1.2~\TeV~and 1.0~\TeV.
Tracks with momenta greater than 1~\GeV~are shown as yellow lines, and energy depositions in the Liquid Argon and Tile calorimeters cells are displayed, respectively, as green and yellow boxes.
}
\label{fig:results:displays}
\end{figure}


\FloatBarrier
\section{Conclusion}
\label{sec:conclusions}

A search for the production of a generic massive resonance $Y$ that decays into two pairs of intermediate resonances $X$, each decaying into two jets, is performed using 140~\ifb\ of proton--proton collisions with $\sqrt{s}=13~\TeV$ collected with the ATLAS Detector during Run 2 of the LHC.
Such a resonant signal in multijet events could be manifested in many models of physics beyond the Standard Model, including in well-motivated models of particle dark matter and models with large extra spatial dimensions.
 
A data-driven background estimate is obtained by fitting these invariant mass distributions with a functional form.
The tetrajet system and average dijet system invariant masses are then studied using the \textsc{BumpHunter} algorithm.
No significant excess of events beyond the Standard Model expectation is observed.
The most significant localized excesses are found at 3200~\GeV in the $\alpha$~region from $0.24 < \alpha < 0.26$ for the~\mfour~spectra (global significance of 0.53 standard deviations), and at 800~\GeV in the $\alpha$~region from $0.26 < \alpha < 0.28$ for the~\mtwoavg~spectra (global significance of 1.98 standard deviations).
The highest tetrajet invariant mass observed is $\mfour=6.6~\TeV$, with a  corresponding~\mtwoavg~value of $2.2~\TeV$.
Using the observed data, upper limits are set on the production cross-sections of new physics scenarios as a function of the $Y$ and $X$ masses in both the model-dependent and model-independent interpretations.
 
Data distributions from this search are openly available on the HEPData platform~\cite{hepdata.140530} for use in future reinterpretations.


\section*{Acknowledgments}


We thank CERN for the very successful operation of the LHC, as well as the
support staff from our institutions without whom ATLAS could not be
operated efficiently.
 
We acknowledge the support of
ANPCyT, Argentina;
YerPhI, Armenia;
ARC, Australia;
BMWFW and FWF, Austria;
ANAS, Azerbaijan;
CNPq and FAPESP, Brazil;
NSERC, NRC and CFI, Canada;
CERN;
ANID, Chile;
CAS, MOST and NSFC, China;
Minciencias, Colombia;
MEYS CR, Czech Republic;
DNRF and DNSRC, Denmark;
IN2P3-CNRS and CEA-DRF/IRFU, France;
SRNSFG, Georgia;
BMBF, HGF and MPG, Germany;
GSRI, Greece;
RGC and Hong Kong SAR, China;
ISF and Benoziyo Center, Israel;
INFN, Italy;
MEXT and JSPS, Japan;
CNRST, Morocco;
NWO, Netherlands;
RCN, Norway;
MEiN, Poland;
FCT, Portugal;
MNE/IFA, Romania;
MESTD, Serbia;
MSSR, Slovakia;
ARRS and MIZ\v{S}, Slovenia;
DSI/NRF, South Africa;
MICINN, Spain;
SRC and Wallenberg Foundation, Sweden;
SERI, SNSF and Cantons of Bern and Geneva, Switzerland;
MOST, Taiwan;
TENMAK, T\"urkiye;
STFC, United Kingdom;
DOE and NSF, United States of America.
In addition, individual groups and members have received support from
BCKDF, CANARIE, Compute Canada and CRC, Canada;
PRIMUS 21/SCI/017 and UNCE SCI/013, Czech Republic;
COST, ERC, ERDF, Horizon 2020, ICSC-NextGenerationEU and Marie Sk{\l}odowska-Curie Actions, European Union;
Investissements d'Avenir Labex, Investissements d'Avenir Idex and ANR, France;
DFG and AvH Foundation, Germany;
Herakleitos, Thales and Aristeia programmes co-financed by EU-ESF and the Greek NSRF, Greece;
BSF-NSF and MINERVA, Israel;
Norwegian Financial Mechanism 2014-2021, Norway;
NCN and NAWA, Poland;
La Caixa Banking Foundation, CERCA Programme Generalitat de Catalunya and PROMETEO and GenT Programmes Generalitat Valenciana, Spain;
G\"{o}ran Gustafssons Stiftelse, Sweden;
The Royal Society and Leverhulme Trust, United Kingdom.
 
The crucial computing support from all WLCG partners is acknowledged gratefully, in particular from CERN, the ATLAS Tier-1 facilities at TRIUMF (Canada), NDGF (Denmark, Norway, Sweden), CC-IN2P3 (France), KIT/GridKA (Germany), INFN-CNAF (Italy), NL-T1 (Netherlands), PIC (Spain), ASGC (Taiwan), RAL (UK) and BNL (USA), the Tier-2 facilities worldwide and large non-WLCG resource providers. Major contributors of computing resources are listed in Ref.~\cite{ATL-SOFT-PUB-2023-001}.


\clearpage
\appendix
 
\printbibliography

@misc{hepdata.140530,
    author = "",
    title = "{Pursuit of paired dijet resonances in the Run 2 dataset with ATLAS}",
    howpublished = "{HEPData (collection)}",
    year = 2023,
    note = "\url{https://doi.org/10.17182/hepdata.140530}"
}

@article{Salam:2010nqg,
    author = "Salam, Gavin P.",
    title = "{Towards Jetography}",
    eprint = "0906.1833",
    archivePrefix = "arXiv",
    primaryClass = "hep-ph",
    doi = "10.1140/epjc/s10052-010-1314-6",
    journal = "Eur. Phys. J. C",
    volume = "67",
    pages = "637--686",
    year = "2010"
}

@article{Webber:1983if,
      author         = "Webber, B. R.",
      title          = "{A QCD model for jet fragmentation including soft gluon
                        interference}",
      journal        = "Nucl. Phys. B",
      volume         = "238",
      year           = "1984",
      pages          = "492-528",
      doi            = "10.1016/0550-3213(84)90333-X",
      reportNumber   = "CERN-TH-3713",
      SLACcitation   = "%%CITATION = NUPHA,B238,492;%%"
}

@inproceedings{Choudalakis:2011qn,
    author = "Choudalakis, Georgios",
    title = "{On hypothesis testing, trials factor, hypertests and the BumpHunter}",
    booktitle = "{Proceedings, PHYSTAT 2011 Workshop on Statistical Issues Related to Discovery Claims in Search Experiments and Unfolding, CERN, Geneva, Switzerland 17-20 January 2011}",
    eprint = "1101.0390",
    archivePrefix = "arXiv",
    primaryClass = "physics.data-an",
    month = "1",
    year = "2011"
}

@Article{Hinshaw:2012aka,
      author         = "Hinshaw, G. and others",
      title          = "{Nine-Year Wilkinson Microwave Anisotropy Probe (WMAP)
                        Observations: Cosmological Parameter Results}",
      collaboration  = "WMAP",
      journal        = "Astrophys. J. Suppl.",
      volume         = "208",
      year           = "2013",
      pages          = "19",
      doi            = "10.1088/0067-0049/208/2/19",
      eprint         = "1212.5226",
      archivePrefix  = "arXiv",
      primaryClass   = "astro-ph.CO",
      SLACcitation   = "%%CITATION = ARXIV:1212.5226;%%"
}

@Article{Akrami:2018vks,
      author         = "Akrami, Y. and others",
      title          = "{Planck 2018 results. I. Overview and the cosmological
                        legacy of Planck}",
      collaboration  = "Planck",
      year           = "2018",
      eprint         = "1807.06205",
      archivePrefix  = "arXiv",
      primaryClass   = "astro-ph.CO",
      SLACcitation   = "%%CITATION = ARXIV:1807.06205;%%"
}

@Article{Feng:2010gw,
      author         = "Feng, Jonathan L.",
      title          = "{Dark Matter Candidates from Particle Physics and Methods
                        of Detection}",
      journal        = "Ann. Rev. Astron. Astrophys.",
      volume         = "48",
      year           = "2010",
      pages          = "495-545",
      doi            = "10.1146/annurev-astro-082708-101659",
      eprint         = "1003.0904",
      archivePrefix  = "arXiv",
      primaryClass   = "astro-ph.CO",
      reportNumber   = "UCI-TR-2009-13",
      SLACcitation   = "%%CITATION = ARXIV:1003.0904;%%"
}

@Article{Bertone:2004pz,
      author         = "Bertone, Gianfranco and Hooper, Dan and Silk, Joseph",
      title          = "{Particle dark matter: Evidence, candidates and
                        constraints}",
      journal        = "Phys. Rept.",
      volume         = "405",
      year           = "2005",
      pages          = "279-390",
      doi            = "10.1016/j.physrep.2004.08.031",
      eprint         = "hep-ph/0404175",
      archivePrefix  = "arXiv",
      primaryClass   = "hep-ph",
      reportNumber   = "FERMILAB-PUB-04-047-A",
      SLACcitation   = "%%CITATION = HEP-PH/0404175;%%"
}

@Article{Trimble1987,
      author         = "Trimble, V.",
      title          = "{Existence and Nature of Dark Matter in the Universe}",
      journal        = "{Ann. Rev. Astron. Astrophys.}",
      volume         = "25",
      year          = "1987",
      pages          = "425-427",
      doi = {10.1146/annurev.aa.25.090187.002233},
}

@Booklet{Arrenberg:2013rzp,
      author         = "Arrenberg, Sebastian and others",
      title          = "{Dark Matter in the Coming Decade: Complementary Paths to Discovery and Beyond}",
      booktitle      = "{Proceedings, 2013 Community Summer Study on the Future
                        of U.S. Particle Physics: Snowmass on the Mississippi
                        (CSS2013): Minneapolis, MN, USA, July 29-August 6, 2013}",
      year           = "2013",
      eprint         = "1310.8621",
      archivePrefix  = "arXiv",
      primaryClass   = "hep-ph",
      reportNumber   = "UFIFT-HEP-13-001, FERMILAB-CONF-13-574-AE",
      SLACcitation   = "%%CITATION = ARXIV:1310.8621;%%"
}

@Article{Roos2012,
      author = "Matts Roos",
      title="{Astrophysical and cosmological probes of dark matter}",
      journal        = " J. of Modern Physics",
      volume         = "3",
      year           = "2012",
      pages          = "1152",
      doi            = "10.4236/jmp.2012.329150",
      eprint         = "1208.3662",
      archivePrefix  = "arXiv",

}

@Article{Aaltonen:2008vt,
      author         = "{CDF Collaboration}",
      title          = "{Global Search for New Physics with 2.0 fb$^{-1}$ at
                        CDF}",
      collaboration  = "CDF",
      journal        = "Phys. Rev. D",
      volume         = "79",
      year           = "2009",
      pages          = "011101",
      doi            = "10.1103/PhysRevD.79.011101",
      eprint         = "0809.3781",
      archivePrefix  = "arXiv",
      primaryClass   = "hep-ex",
      reportNumber   = "FERMILAB-PUB-08-400-E",
      SLACcitation   = "%%CITATION = ARXIV:0809.3781;%%"
}

@article{Boelaert:2010inn,
    author = "Boelaert, Nele and Akesson, Torsten",
    title = "{Dijet angular distributions at $\sqrt{s} =14$ TeV}",
    eprint = "0905.3961",
    archivePrefix = "arXiv",
    primaryClass = "hep-ph",
    doi = "10.1140/epjc/s10052-010-1268-8",
    journal = "Eur. Phys. J. C",
    volume = "66",
    pages = "343--357",
    year = "2010"
}

@article{Gingrich:2009hj,
    author = "Gingrich, Douglas M.",
    title = "{Quantum black holes with charge, colour, and spin at the LHC}",
    eprint = "0912.0826",
    archivePrefix = "arXiv",
    primaryClass = "hep-ph",
    doi = "10.1088/0954-3899/37/10/105008",
    journal = "J. Phys. G",
    volume = "37",
    pages = "105008",
    year = "2010"
}

@article{Calmet:2008dg,
    author = "Calmet, Xavier and Gong, Wei and Hsu, Stephen D. H.",
    title = "{Colorful quantum black holes at the LHC}",
    eprint = "0806.4605",
    archivePrefix = "arXiv",
    primaryClass = "hep-ph",
    reportNumber = "CP3-08-22",
    doi = "10.1016/j.physletb.2008.08.011",
    journal = "Phys. Lett. B",
    volume = "668",
    pages = "20--23",
    year = "2008"
}

@article{Giddings:2001bu,
    author = "Giddings, Steven B. and Thomas, Scott D.",
    title = "{High energy colliders as black hole factories: The end of short distance physics}",
    eprint = "hep-ph/0106219",
    archivePrefix = "arXiv",
    reportNumber = "NSF-ITP-01-62, SU-ITP-01-30",
    doi = "10.1103/PhysRevD.65.056010",
    journal = "Phys. Rev. D",
    volume = "65",
    pages = "056010",
    year = "2002"
}

@article{Dimopoulos:2001hw,
    author = "Dimopoulos, Savas and Landsberg, Greg L.",
    title = "{Black Holes at the Large Hadron Collider}",
    eprint = "hep-ph/0106295",
    archivePrefix = "arXiv",
    reportNumber = "SU-ITP-01-31",
    doi = "10.1103/PhysRevLett.87.161602",
    journal = "Phys. Rev. Lett.",
    volume = "87",
    pages = "161602",
    year = "2001"
}

@article{Meade:2007sz,
    author = "Meade, Patrick and Randall, Lisa",
    title = "{Black Holes and Quantum Gravity at the LHC}",
    eprint = "0708.3017",
    archivePrefix = "arXiv",
    primaryClass = "hep-ph",
    doi = "10.1088/1126-6708/2008/05/003",
    journal = "JHEP",
    volume = "05",
    pages = "003",
    year = "2008"
}

@article{Anchordoqui:2003ug,
    author = "Anchordoqui, Luis A. and Feng, Jonathan L. and Goldberg, Haim and Shapere, Alfred D.",
    title = "{Inelastic black hole production and large extra dimensions}",
    eprint = "hep-ph/0311365",
    archivePrefix = "arXiv",
    reportNumber = "NUB-3243-TH-03, UCI-TR-2003-30, UK-03-15",
    doi = "10.1016/j.physletb.2004.05.051",
    journal = "Phys. Lett. B",
    volume = "594",
    pages = "363--367",
    year = "2004"
}

@article{UA1:1988pgb,
    author = "{UA1 Collaboration}",
    collaboration = "UA1",
    title = "{Two-jet mass distributions at the CERN proton - antiproton collider}",
    reportNumber = "CERN-EP-88-54",
    doi = "10.1016/0370-2693(88)91843-6",
    journal = "Phys. Lett. B",
    volume = "209",
    pages = "127--134",
    year = "1988"
}

@article{UA2:1993tee,
    author = "{UA2 Collaboration}",
    title = "{A Search for new intermediate vector mesons and excited quarks decaying to two jets at the CERN $\bar{p} p$ collider}",
    reportNumber = "CERN-PPE-93-66",
    doi = "10.1016/0550-3213(93)90395-6",
    journal = "Nucl. Phys. B",
    volume = "400",
    pages = "3--24",
    year = "1993"
}

@article{CDF:2008ieg,
    author = "{CDF Collaboration}",
    title = "{Search for new particles decaying into dijets in proton-antiproton collisions at $\sqrt{s} = 1.96$ TeV}",
    eprint = "0812.4036",
    archivePrefix = "arXiv",
    primaryClass = "hep-ex",
    reportNumber = "FERMILAB-PUB-08-572-E",
    doi = "10.1103/PhysRevD.79.112002",
    journal = "Phys. Rev. D",
    volume = "79",
    pages = "112002",
    year = "2009"
}

@article{D0:2009rxw,
    author = "{D0 Collaboration}",
    title = "{Measurement of dijet angular distributions at $\sqrt{s} = 1.96$ TeV and searches for quark compositeness and extra spatial dimensions}",
    eprint = "0906.4819",
    archivePrefix = "arXiv",
    primaryClass = "hep-ex",
    reportNumber = "FERMILAB-PUB-09-326-E",
    doi = "10.1103/PhysRevLett.103.191803",
    journal = "Phys. Rev. Lett.",
    volume = "103",
    pages = "191803",
    year = "2009"
}

@article{LUCID2,
  author={G. Avoni and others},
  title={The new LUCID-2 detector for luminosity measurement and monitoring in ATLAS},
  journal={JINST},
  volume={13},
  number={07},
  pages={P07017},
  doi="10.1088/1748-0221/13/07/P07017",
  year={2018},
}

@software{pyBumpHunter,
  author = "Vaslin, Louis and Donini, Julien and Eschle, Jonas and Duarte, Javier",
  title = "{pyBumpHunter: v0.4.1}",
  version = {0.4.1},
  doi = {10.5281/zenodo.7684559},
  url = {https://doi.org/10.5281/zenodo.7684558},
  note = {\url{https://github.com/scikit-hep/pyBumpHunter/releases/tag/v0.4.1}},
  year={2023},
  month={Feb}
}

@article{Vaslin:2022bds,
    author = "Vaslin, Louis and Calvet, Samuel and Barra, Vincent and Donini, Julien",
    title = "{pyBumpHunter: A model independent bump hunting tool in Python for High Energy Physics analyses}",
    eprint = "2208.14760",
    archivePrefix = "arXiv",
    primaryClass = "hep-ex",
    month = "8",
    year = "2022"
}

@article{Adragna:2009zz,
    author = "Adragna, P. and others",
    title = "{Testbeam studies of production modules of the ATLAS tile calorimeter}",
    reportNumber = "ATL-TILECAL-PUB-2009-002, ATL-COM-TILECAL-2009-004",
    doi = "10.1016/j.nima.2009.04.009",
    journal = "Nucl. Instrum. Meth. A",
    volume = "606",
    pages = "362--394",
    year = "2009"
}

@article{ATLASLiquidArgonHEC:2002hwj,
    author = "Dowler, B and others",
    collaboration = "ATLAS Liquid Argon HEC",
    title = "{Performance of the ATLAS hadronic end-cap calorimeter in beam tests}",
    doi = "10.1016/S0168-9002(01)01338-9",
    journal = "Nucl. Instrum. Meth. A",
    volume = "482",
    pages = "94--124",
    year = "2002"
}

@phdthesis{Choudalakis:2008pr,
    author = "Choudalakis, Georgios",
    title = "{Model Independent Search For New Physics At The Tevatron}",
    eprint = "0805.3954",
    archivePrefix = "arXiv",
    primaryClass = "hep-ex",
    reportNumber = "FERMILAB-THESIS-2008-10",
    doi = "10.2172/929119",
    school = "MIT",
    year = "2008"
}

@mastersthesis{Oreglia:1980cs,
    author = "Oreglia, M.",
    title = "{A Study of the Reactions $\psi^\prime \to \gamma \gamma \psi$}",
    reportNumber = "SLAC-0236, SLAC-236, UMI-81-08973, SLAC-R-0236, SLAC-R-236",
    type = "PhD thesis",
    school = "Stanford University",
    month = "12",
    year = "1980",
    note    = "SLAC-R-0236",
}

@article{Assad:2017iib,
    author = "Assad, Nima and Fornal, Bartosz and Grinstein, Benjamin",
    title = "{Baryon Number and Lepton Universality Violation in Leptoquark and Diquark Models}",
    eprint = "1708.06350",
    archivePrefix = "arXiv",
    primaryClass = "hep-ph",
    doi = "10.1016/j.physletb.2017.12.042",
    journal = "Phys. Lett. B",
    volume = "777",
    pages = "324--331",
    year = "2018"
}

@article{Mohapatra:2007af,
    author = "Mohapatra, R. N. and Okada, Nobuchika and Yu, Hai-Bo",
    title = "{Diquark Higgs at LHC}",
    eprint = "0709.1486",
    archivePrefix = "arXiv",
    primaryClass = "hep-ph",
    reportNumber = "UMD-PP-07-010",
    doi = "10.1103/PhysRevD.77.011701",
    journal = "Phys. Rev. D",
    volume = "77",
    pages = "011701",
    year = "2008"
}

@article{Han:2009ya,
    author = "Han, Tao and Lewis, Ian and McElmurry, Thomas",
    title = "{QCD Corrections to Scalar Diquark Production at Hadron Colliders}",
    eprint = "0909.2666",
    archivePrefix = "arXiv",
    primaryClass = "hep-ph",
    reportNumber = "MADPH-09-1536",
    doi = "10.1007/JHEP01(2010)123",
    journal = "JHEP",
    volume = "01",
    pages = "123",
    year = "2010"
}

@article{Richardson:2011df,
    author = "Richardson, Peter and Winn, David",
    title = "{Simulation of Sextet Diquark Production}",
    eprint = "1108.6154",
    archivePrefix = "arXiv",
    primaryClass = "hep-ph",
    reportNumber = "IPPP-11-50, DCPT-11-100",
    doi = "10.1140/epjc/s10052-012-1862-z",
    journal = "Eur. Phys. J. C",
    volume = "72",
    pages = "1862",
    year = "2012"
}

@article{Chivukula:1996yr,
    author = "Chivukula, R. S. and Cohen, Andrew G. and Simmons, Elizabeth H.",
    title = "{New strong interactions at the Tevatron?}",
    eprint = "hep-ph/9603311",
    archivePrefix = "arXiv",
    reportNumber = "BUHEP-96-5",
    doi = "10.1016/0370-2693(96)00464-9",
    journal = "Phys. Lett. B",
    volume = "380",
    pages = "92--98",
    year = "1996"
}

@article{Bai:2010dj,
    author = "Bai, Yang and Dobrescu, Bogdan A.",
    title = "{Heavy Octets and Tevatron Signals with Three or Four b Jets}",
    eprint = "1012.5814",
    archivePrefix = "arXiv",
    primaryClass = "hep-ph",
    reportNumber = "SLAC-PUB-14343, FERMILAB-PUB-10-534-T",
    doi = "10.1007/JHEP07(2011)100",
    journal = "JHEP",
    volume = "07",
    pages = "100",
    year = "2011"
}

@article{Bai:2018jsr,
    author = "Bai, Yang and Dobrescu, Bogdan A.",
    title = "{Collider Tests of the Renormalizable Coloron Model}",
    eprint = "1802.03005",
    archivePrefix = "arXiv",
    primaryClass = "hep-ph",
    reportNumber = "FERMILAB-PUB-17-555-T",
    doi = "10.1007/JHEP04(2018)114",
    journal = "JHEP",
    volume = "04",
    pages = "114",
    year = "2018"
}

@article{Chivukula:2013xka,
    author = "Chivukula, R. Sekhar and Farzinnia, Arsham and Ren, Jing and Simmons, Elizabeth H.",
    title = "{Constraints on the Scalar Sector of the Renormalizable Coloron Model}",
    eprint = "1307.1064",
    archivePrefix = "arXiv",
    primaryClass = "hep-ph",
    reportNumber = "MSUHEP-130703",
    doi = "10.1103/PhysRevD.88.075020",
    journal = "Phys. Rev. D",
    volume = "88",
    number = "7",
    pages = "075020",
    year = "2013",
    note = "[Erratum: \href{https://link.aps.org/doi/10.1103/PhysRevD.89.059905}{Phys. Rev. D \textbf{89} (2014) 059905}]"
}

@article{Dobrescu:2018psr,
    author = "Dobrescu, Bogdan A. and Harris, Robert M. and Isaacson, Joshua",
    title = "{Ultraheavy resonances at the LHC: beyond the QCD background}",
    eprint = "1810.09429",
    archivePrefix = "arXiv",
    primaryClass = "hep-ph",
    reportNumber = "FERMILAB-PUB-18-561-T",
    month = "10",
    year = "2018"
}

@article{Kilic:2008pm,
    author = "Kilic, Can and Okui, Takemichi and Sundrum, Raman",
    title = "{Colored Resonances at the Tevatron: Phenomenology and Discovery Potential in Multijets}",
    eprint = "0802.2568",
    archivePrefix = "arXiv",
    primaryClass = "hep-ph",
    reportNumber = "UMD-PP-08-002",
    doi = "10.1088/1126-6708/2008/07/038",
    journal = "JHEP",
    volume = "07",
    pages = "038",
    year = "2008"
}

@article{Bai:2011mr,
    author = "Bai, Yang and Shelton, Jessie",
    title = "{Composite Octet Searches with Jet Substructure}",
    eprint = "1107.3563",
    archivePrefix = "arXiv",
    primaryClass = "hep-ph",
    reportNumber = "SLAC-PUB-14509",
    doi = "10.1007/JHEP07(2012)067",
    journal = "JHEP",
    volume = "07",
    pages = "067",
    year = "2012"
}

@Article{Cacciari:2008gp,
     author    = "Cacciari, Matteo and Salam, Gavin P. and Soyez, Gregory",
     title     = "{The anti-\(k_{t}\) jet clustering algorithm}",
     journal   = "JHEP",
     volume    = "04",
     year      = "2008",
     pages     = "063",
     eprint    = "0802.1189",
     archivePrefix = "arXiv",
     primaryClass  =  "hep-ph",
     doi       = "10.1088/1126-6708/2008/04/063",
     SLACcitation  = "%%CITATION = 0802.1189;%%"
}

@Article{Fastjet,
      author         = "Cacciari, Matteo and Salam, Gavin P. and Soyez, Gregory",
      title          = "{FastJet user manual}",
      journal        = "Eur. Phys. J. C",
      volume         = "72",
      year           = "2012",
      pages          = "1896",
      doi            = "10.1140/epjc/s10052-012-1896-2",
      eprint         = "1111.6097",
      archivePrefix  = "arXiv",
      primaryClass   = "hep-ph",
      reportNumber   = "CERN-PH-TH-2011-297",
      SLACcitation   = "%%CITATION = ARXIV:1111.6097;%%"
}

@article{Lai:2010vv,
      author         = "Lai, H.-L. and others",
      title          = "{New parton distributions for collider physics}",
      journal        = "Phys. Rev. D",
      volume         = "82",
      pages          = "074024",
      doi            = "10.1103/PhysRevD.82.074024",
      year           = "2010",
      eprint         = "1007.2241",
      archivePrefix  = "arXiv",
      primaryClass   = "hep-ph",
      reportNumber   = "MSUHEP-100707, SMU-HEP-10-10",
      SLACcitation   = "%%CITATION = ARXIV:1007.2241;%%",
}

@article{Dulat:2015mca,
      author         = "Dulat, Sayipjamal and Hou, Tie-Jiun and Gao, Jun and
                        Guzzi, Marco and Huston, Joey and Nadolsky, Pavel and
                        Pumplin, Jon and Schmidt, Carl and Stump, Daniel and Yuan,
                        C. P.",
      title          = "{New parton distribution functions from a global analysis
                        of quantum chromodynamics}",
      journal        = "Phys. Rev. D",
      volume         = "93",
      year           = "2016",
      number         = "3",
      pages          = "033006",
      doi            = "10.1103/PhysRevD.93.033006",
      eprint         = "1506.07443",
      archivePrefix  = "arXiv",
      primaryClass   = "hep-ph",
      SLACcitation   = "%%CITATION = ARXIV:1506.07443;%%"
}

@article{Ball:2012cx,
      author         = "{NNPDF Collaboration} and Ball, Richard D. and others",
      title          = "{Parton distributions with LHC data}",
      journal        = "Nucl. Phys. B",
      volume         = "867",
      year           = "2013",
      pages          = "244",
      doi            = "10.1016/j.nuclphysb.2012.10.003",
      eprint         = "1207.1303",
      archivePrefix  = "arXiv",
      primaryClass   = "hep-ph",
      reportNumber   = "EDINBURGH-2012-08, IFUM-FT-997, FR-PHENO-2012-014,
                        RWTH-TTK-12-25, CERN-PH-TH-2012-037, SFB-CPP-12-47\,
                        --CERN-PH-TH-2012-037",
      SLACcitation   = "%%CITATION = ARXIV:1207.1303;%%"
}

@Article{Sjostrand:2006za,
     author    = "Sj{\"o}strand, Torbjorn and Mrenna, Stephen and Skands, Peter Z.",
     title     = "{PYTHIA 6.4 physics and manual}",
     journal   = "JHEP",
     volume    = "05",
     year      = "2006",
     pages     = "026",
     eprint    = "hep-ph/0603175",
     archivePrefix = "arXiv",
     doi       = "10.1088/1126-6708/2006/05/026",
     SLACcitation  = "%%CITATION = HEP-PH/0603175;%%"
}

@article{Mrenna:2016sih,
     author         = "Mrenna, S. and Skands, P.",
     title          = "{Automated parton-shower variations in PYTHIA 8}",
     journal        = "Phys. Rev. D",
     volume         = "94",
     year           = "2016",
     pages          = "074005",
     doi            = "10.1103/PhysRevD.94.074005",
     eprint         = "1605.08352",
     archivePrefix  = "arXiv",
     primaryClass   = "hep-ph",
}

@article{Sjostrand:2014zea,
      author         = "Sj{\"o}strand, Torbj{\"o}rn and Ask, Stefan and Christiansen,
                        Jesper R. and Corke, Richard and Desai, Nishita and Ilten,
                        Philip and Mrenna, Stephen and Prestel, Stefan and
                        Rasmussen, Christine O. and Skands, Peter Z.",
      title          = "{An introduction to PYTHIA 8.2}",
      journal        = "Comput. Phys. Commun.",
      volume         = "191",
      year           = "2015",
      pages          = "159",
      doi            = "10.1016/j.cpc.2015.01.024",
      eprint         = "1410.3012",
      archivePrefix  = "arXiv",
      primaryClass   = "hep-ph",
      reportNumber   = "LU-TP-14-36, MCNET-14-22, CERN-PH-TH-2014-190,
                        FERMILAB-PUB-14-316-CD, DESY-14-178, SLAC-PUB-16122,
                        --FERMILAB-PUB-14-316-CD",
      SLACcitation   = "%%CITATION = ARXIV:1410.3012;%%"
}

@Article{Lange:2001uf,
      author         = "Lange, D. J.",
      title          = "{The EvtGen particle decay simulation package}",
      booktitle      = "{Proceedings, 7th International Conference on B physics
                        at hadron machines (BEAUTY 2000)}",
      journal        = "Nucl. Instrum. Meth. A",
      volume         = "462",
      year           = "2001",
      pages          = "152",
      doi            = "10.1016/S0168-9002(01)00089-4",
      SLACcitation   = "%%CITATION = NUIMA,A462,152;%%"
}

@Article{Gleisberg:2008ta,
      author         = "Gleisberg, T. and H{\"o}che, S. and Krauss, F. and Sch{\"o}nherr, M. and Schumann, S. and Siegert, F. and Winter J.",
      title          = "{Event generation with SHERPA 1.1}",
      journal        = "JHEP",
      volume         = "02",
      pages          = "007",
      doi            = "10.1088/1126-6708/2009/02/007",
      year           = "2009",
      eprint         = "0811.4622",
      archivePrefix  = "arXiv",
      primaryClass   = "hep-ph",
}

@Article{Schumann:2007mg,
      author         = "Schumann, Steffen and Krauss, Frank",
      title          = "{A parton shower algorithm based on Catani--Seymour dipole factorisation}",
      journal        = "JHEP",
      volume         = "03",
      pages          = "038",
      doi            = "10.1088/1126-6708/2008/03/038",
      year           = "2008",
      eprint         = "0709.1027",
      archivePrefix  = "arXiv",
      primaryClass   = "hep-ph",
}

@article{Agostinelli:2002hh,
      author         = "Agostinelli, S. and others",
      title          = "{\textsc{Geant4} -- a simulation toolkit}",
      journal        = "Nucl. Instrum. Meth. A",
      volume         = "506",
      year           = "2003",
      pages          = "250",
      doi            = "10.1016/S0168-9002(03)01368-8",
      reportNumber   = "SLAC-PUB-9350, FERMILAB-PUB-03-339",
      SLACcitation   = "%%CITATION = NUIMA,A506,250;%%"
}

@Article{Read:2002hq,
    Author = {Read, Alexander L.},
    Journal = {J. Phys. G},
    Pages = {2693},
    doi = "10.1088/0954-3899/28/10/313",
    Title = "{Presentation of search results: the \(CL_S\) technique}",
    Volume = {28},
    Year = {2002}
}

@Article{Cowan:2010js,
   Author = {Cowan, Glen and Cranmer, Kyle and Gross, Eilam and Vitells, Ofer},
   Title = {{Asymptotic formulae for likelihood-based tests of new physics}},
   Journal = {Eur. Phys. J. C},
   Volume = {71},
   Year = {2011},
   Pages = {1554},
   doi = "10.1140/epjc/s10052-011-1554-0",
   Eprint = "1007.1727",
   Archiveprefix = {arXiv},
   Primaryclass = {physics.data-an},
   related = "Cowan:2010js-err",
   relatedstring  = "Erratum:",
}

@Article{PERF-2007-01,
    author         = "{ATLAS Collaboration}",
    title          = "{The ATLAS Experiment at the CERN Large Hadron Collider}",
    journal        = "JINST",
    volume         = "3",
    year           = "2008",
    pages          = "S08003",
    doi            = "10.1088/1748-0221/3/08/S08003",
    primaryClass   = "hep-ex",
}

@Article{EXOT-2010-01,
    author         = "{ATLAS Collaboration}",
    title          = "{Search for New Particles in Two-Jet Final States in \(7\,\text{TeV}\) Proton--Proton Collisions with the ATLAS Detector at the LHC}",
    journal        = "Phys. Rev. Lett.",
    volume         = "105",
    year           = "2010",
    pages          = "161801",
    doi            = "10.1103/PhysRevLett.105.161801",
    eprint         = "1008.2461",
    archivePrefix  = "arXiv",
    primaryClass   = "hep-ex",
}

@Article{EXOT-2010-07,
    author         = "{ATLAS Collaboration}",
    title          = "{A search for new physics in dijet mass and angular distributions in \(pp\) collisions at \(\sqrt{s} = 7\,\text{TeV}\) measured with the ATLAS detector}",
    journal        = "New J. Phys.",
    volume         = "13",
    year           = "2011",
    pages          = "053044",
    doi            = "10.1088/1367-2630/13/5/053044",
    reportNumber   = "CERN-PH-EP-2011-030",
    eprint         = "1103.3864",
    archivePrefix  = "arXiv",
    primaryClass   = "hep-ex",
}

@Article{SOFT-2010-01,
    author         = "{ATLAS Collaboration}",
    title          = "{The ATLAS Simulation Infrastructure}",
    journal        = "Eur. Phys. J. C",
    volume         = "70",
    year           = "2010",
    pages          = "823",
    doi            = "10.1140/epjc/s10052-010-1429-9",
    eprint         = "1005.4568",
    archivePrefix  = "arXiv",
    primaryClass   = "physics.ins-det",
}

@Article{EXOT-2011-07,
    author         = "{ATLAS Collaboration}",
    title          = "{Search for new physics in the dijet mass distribution using \(1\,\text{fb}^{-1}\) of \(pp\) collision data at \(\sqrt{s} = 7\,\text{TeV}\) collected by the ATLAS detector}",
    journal        = "Phys. Lett. B",
    volume         = "708",
    year           = "2012",
    pages          = "37",
    doi            = "10.1016/j.physletb.2012.01.035",
    reportNumber   = "CERN-PH-EP-2011-127",
    eprint         = "1108.6311",
    archivePrefix  = "arXiv",
    primaryClass   = "hep-ex",
}

@Article{EXOT-2011-21,
    author         = "{ATLAS Collaboration}",
    title          = "{ATLAS search for new phenomena in dijet mass and angular distributions using \(pp\) collisions at \(\sqrt{s} = 7\,\text{TeV}\)}",
    journal        = "JHEP",
    volume         = "01",
    year           = "2013",
    pages          = "029",
    doi            = "10.1007/JHEP01(2013)029",
    reportNumber   = "CERN-PH-EP-2012-257",
    eprint         = "1210.1718",
    archivePrefix  = "arXiv",
    primaryClass   = "hep-ex",
}

@Article{SUSY-2012-18,
    author         = "{ATLAS Collaboration}",
    title          = "{Search for pair-produced massive coloured scalars in four-jet final states with the ATLAS detector in proton--proton collisions at \(\sqrt{s} = 7\,\text{TeV}\)}",
    journal        = "Eur. Phys. J. C",
    volume         = "73",
    year           = "2013",
    pages          = "2263",
    doi            = "10.1140/epjc/s10052-012-2263-z",
    eprint         = "1210.4826",
    archivePrefix  = "arXiv",
    primaryClass   = "hep-ex",
}

@Article{EXOT-2013-11,
    author         = "{ATLAS Collaboration}",
    title          = "{Search for new phenomena in the dijet mass distribution using \(pp\) collision data at \(\sqrt{s} = 8\,\text{TeV}\) with the ATLAS detector}",
    journal        = "Phys. Rev. D",
    volume         = "91",
    year           = "2015",
    pages          = "052007",
    doi            = "10.1103/PhysRevD.91.052007",
    reportNumber   = "CERN-PH-EP-2014-147",
    eprint         = "1407.1376",
    archivePrefix  = "arXiv",
    primaryClass   = "hep-ex",
}

@Article{PERF-2014-07,
    author         = "{ATLAS Collaboration}",
    title          = "{Topological cell clustering in the ATLAS calorimeters and its performance in LHC Run~1}",
    journal        = "Eur. Phys. J. C",
    volume         = "77",
    year           = "2017",
    pages          = "490",
    doi            = "10.1140/epjc/s10052-017-5004-5",
    reportNumber   = "CERN-PH-EP-2015-304",
    eprint         = "1603.02934",
    archivePrefix  = "arXiv",
    primaryClass   = "hep-ex",
}

@Article{EXOT-2015-02,
    author         = "{ATLAS Collaboration}",
    title          = "{Search for new phenomena in dijet mass and angular distributions from \(pp\) collisions at \(\sqrt{s} = 13\,\text{TeV}\) with the ATLAS detector}",
    journal        = "Phys. Lett. B",
    volume         = "754",
    year           = "2016",
    pages          = "302",
    doi            = "10.1016/j.physletb.2016.01.032",
    reportNumber   = "CERN-PH-EP-2015-311",
    eprint         = "1512.01530",
    archivePrefix  = "arXiv",
    primaryClass   = "hep-ex",
}

@Article{PERF-2015-05,
    author         = "{ATLAS Collaboration}",
    title          = "{A measurement of the calorimeter response to single hadrons and determination of the jet energy scale uncertainty using LHC Run-1 \(pp\)-collision data with the ATLAS detector}",
    journal        = "Eur. Phys. J. C",
    volume         = "77",
    year           = "2017",
    pages          = "26",
    doi            = "10.1140/epjc/s10052-016-4580-0",
    reportNumber   = "CERN-EP-2016-149",
    eprint         = "1607.08842",
    archivePrefix  = "arXiv",
    primaryClass   = "hep-ex",
}

@Article{PERF-2015-09,
    author         = "{ATLAS Collaboration}",
    title          = "{Jet reconstruction and performance using particle flow with the ATLAS Detector}",
    journal        = "Eur. Phys. J. C",
    volume         = "77",
    year           = "2017",
    pages          = "466",
    doi            = "10.1140/epjc/s10052-017-5031-2",
    reportNumber   = "CERN-EP-2017-024",
    eprint         = "1703.10485",
    archivePrefix  = "arXiv",
    primaryClass   = "hep-ex",
}

@Article{SUSY-2015-05,
    author         = "{ATLAS Collaboration}",
    title          = "{A search for top squarks with \(R\)-parity-violating decays to all-hadronic final states with the ATLAS detector in \(\sqrt{s} = 8\,\text{TeV}\) proton--proton collisions}",
    journal        = "JHEP",
    volume         = "06",
    year           = "2016",
    pages          = "067",
    doi            = "10.1007/JHEP06(2016)067",
    reportNumber   = "CERN-PH-EP-2015-296",
    eprint         = "1601.07453",
    archivePrefix  = "arXiv",
    primaryClass   = "hep-ex",
}

@Article{EXOT-2016-21,
    author         = "{ATLAS Collaboration}",
    title          = "{Search for new phenomena in dijet events using \(37\,\text{fb}^{-1}\) of \(pp\) collision data collected at \(\sqrt{s} = 13\,\text{TeV}\) with the ATLAS detector}",
    journal        = "Phys. Rev. D",
    volume         = "96",
    year           = "2017",
    pages          = "052004",
    doi            = "10.1103/PhysRevD.96.052004",
    reportNumber   = "CERN-EP-2017-042",
    eprint         = "1703.09127",
    archivePrefix  = "arXiv",
    primaryClass   = "hep-ex",
}

@Article{PERF-2016-07,
    author         = "{ATLAS Collaboration}",
    title          = "{Performance of missing transverse momentum reconstruction with the ATLAS detector using proton--proton collisions at \(\sqrt{s} = 13\,\text{TeV}\)}",
    journal        = "Eur. Phys. J. C",
    volume         = "78",
    year           = "2018",
    pages          = "903",
    doi            = "10.1140/epjc/s10052-018-6288-9",
    reportNumber   = "CERN-EP-2017-274",
    eprint         = "1802.08168",
    archivePrefix  = "arXiv",
    primaryClass   = "hep-ex",
}

@Article{SUSY-2016-09,
    author         = "{ATLAS Collaboration}",
    title          = "{A search for pair-produced resonances in four-jet final states at \(\sqrt{s} = 13\,\text{TeV}\) with the ATLAS detector}",
    journal        = "Eur. Phys. J. C",
    volume         = "78",
    year           = "2018",
    pages          = "250",
    doi            = "10.1140/epjc/s10052-018-5693-4",
    reportNumber   = "CERN-EP-2017-183",
    eprint         = "1710.07171",
    archivePrefix  = "arXiv",
    primaryClass   = "hep-ex",
}

@Article{TRIG-2016-01,
    author         = "{ATLAS Collaboration}",
    title          = "{Performance of the ATLAS trigger system in 2015}",
    journal        = "Eur. Phys. J. C",
    volume         = "77",
    year           = "2017",
    pages          = "317",
    doi            = "10.1140/epjc/s10052-017-4852-3",
    reportNumber   = "CERN-EP-2016-241",
    eprint         = "1611.09661",
    archivePrefix  = "arXiv",
    primaryClass   = "hep-ex",
}

@Article{EXOT-2017-32,
    author         = "{ATLAS Collaboration}",
    title          = "{Constraints on mediator-based dark matter and scalar dark energy models using \(\sqrt{s} = 13\,\text{TeV}\) \(pp\) collision data collected by the ATLAS detector}",
    journal        = "JHEP",
    volume         = "05",
    year           = "2019",
    pages          = "142",
    doi            = "10.1007/JHEP05(2019)142",
    reportNumber   = "CERN-EP-2018-334",
    eprint         = "1903.01400",
    archivePrefix  = "arXiv",
    primaryClass   = "hep-ex",
}

@Article{DAPR-2018-01,
    author         = "{ATLAS Collaboration}",
    title          = "{ATLAS data quality operations and performance for 2015--2018 data-taking}",
    journal        = "JINST",
    volume         = "15",
    year           = "2020",
    pages          = "P04003",
    doi            = "10.1088/1748-0221/15/04/P04003",
    reportNumber   = "CERN-EP-2019-207",
    eprint         = "1911.04632",
    archivePrefix  = "arXiv",
    primaryClass   = "physics.ins-det",
}

@Article{HDBS-2018-59,
    author         = "{ATLAS Collaboration}",
    title          = "{Dijet Resonance Search with Weak Supervision Using \(\sqrt{s} = 13\,\text{TeV}\) \(pp\) collisions in the ATLAS detector}",
    journal        = "Phys. Rev. Lett.",
    volume         = "125",
    year           = "2020",
    pages          = "131801",
    doi            = "10.1103/PhysRevLett.125.131801",
    reportNumber   = "CERN-EP-2020-062",
    eprint         = "2005.02983",
    archivePrefix  = "arXiv",
    primaryClass   = "hep-ex",
}

@Article{JETM-2018-05,
    author         = "{ATLAS Collaboration}",
    title          = "{Jet energy scale and resolution measured in proton--proton collisions at \(\sqrt{s} = 13\,\text{TeV}\) with the ATLAS detector}",
    journal        = "Eur. Phys. J. C",
    volume         = "81",
    year           = "2020",
    pages          = "689",
    doi            = "10.1140/epjc/s10052-021-09402-3",
    reportNumber   = "CERN-EP-2020-083",
    eprint         = "2007.02645",
    archivePrefix  = "arXiv",
    primaryClass   = "hep-ex",
}

@Article{JETM-2018-06,
    author         = "{ATLAS Collaboration}",
    title          = "{Optimisation of large-radius jet reconstruction for the ATLAS detector in \(13\,\text{TeV}\) proton--proton collisions}",
    journal        = "Eur. Phys. J. C",
    volume         = "81",
    year           = "2020",
    pages          = "334",
    doi            = "10.1140/epjc/s10052-021-09054-3",
    reportNumber   = "CERN-EP-2020-134",
    eprint         = "2009.04986",
    archivePrefix  = "arXiv",
    primaryClass   = "hep-ex",
}

@Article{JETM-2018-08,
    author         = "{ATLAS Collaboration}",
    title          = "{Measurement of the energy response of the ATLAS calorimeter to charged pions from \(W^{\pm}\to\tau^{\pm}(\to\pi^{\pm}\nu_{\tau})\nu_{\tau}\) events in Run~2 data}",
    journal        = "Eur. Phys. J. C",
    volume         = "82",
    year           = "2021",
    pages          = "223",
    doi            = "10.1140/epjc/s10052-022-10117-2",
    reportNumber   = "CERN-EP-2021-147",
    eprint         = "2108.09043",
    archivePrefix  = "arXiv",
    primaryClass   = "hep-ex",
}

@Article{PIX-2018-001,
    author         = "Abbott, B. and others",
    title          = "{Production and integration of the ATLAS Insertable B-Layer}",
    journal        = "JINST",
    volume         = "13",
    year           = "2018",
    pages          = "T05008",
    doi            = "10.1088/1748-0221/13/05/T05008",
    eprint         = "1803.00844",
    archivePrefix  = "arXiv",
    primaryClass   = "physics.ins-det",
}

@Article{EXOT-2019-03,
    author         = "{ATLAS Collaboration}",
    title          = "{Search for new resonances in mass distributions of jet pairs using \(139\,\text{fb}^{-1}\) of \(pp\) collisions at \(\sqrt{s} = 13\,\text{TeV}\) with the ATLAS detector}",
    journal        = "JHEP",
    volume         = "03",
    year           = "2020",
    pages          = "145",
    doi            = "10.1007/JHEP03(2020)145",
    reportNumber   = "CERN-EP-2019-162",
    eprint         = "1910.08447",
    archivePrefix  = "arXiv",
    primaryClass   = "hep-ex",
}

@Article{EXOT-2020-15,
    author         = "{ATLAS Collaboration}",
    title          = "{Search for new phenomena in multi-body invariant masses in events with at least one isolated lepton and two jets using \(\sqrt{s} = 13\,\text{TeV}\) proton--proton collision data collected by the ATLAS detector}",
    journal        = "JHEP",
    volume         = "07",
    year           = "2022",
    pages          = "202",
    doi            = "10.1007/JHEP07(2023)202",
    reportNumber   = "CERN-EP-2022-179",
    eprint         = "2211.08945",
    archivePrefix  = "arXiv",
    primaryClass   = "hep-ex",
}

@Article{DAPR-2021-01,
    author         = "{ATLAS Collaboration}",
    title          = "{Luminosity determination in \(pp\) collisions at \(\sqrt{s} = 13\,\text{TeV}\) using the ATLAS detector at the LHC}",
    year           = "2022",
    reportNumber   = "CERN-EP-2022-281",
    eprint         = "2212.09379",
    archivePrefix  = "arXiv",
    primaryClass   = "hep-ex",
}

@Article{EXOT-2022-18,
    author         = "{ATLAS Collaboration}",
    title          = "{Pursuit of paired dijet resonances in the Run~2 dataset with ATLAS}",
    year           = "2023",
    eprint         = "2307.14944",
    archivePrefix  = "arXiv",
    primaryClass   = "hep-ex",
}

@Booklet{ATL-SOFT-PUB-2021-001,
    author         = "{ATLAS Collaboration}",
    title          = "{The ATLAS Collaboration Software and Firmware}",
    howpublished   = "{ATL-SOFT-PUB-2021-001}",
    url            = "https://cds.cern.ch/record/2767187",
    year           = "2021",
}

@Booklet{ATL-SOFT-PUB-2023-001,
    author         = "{ATLAS Collaboration}",
    title          = "{ATLAS Computing Acknowledgements}",
    howpublished   = "{ATL-SOFT-PUB-2023-001}",
    url            = "https://cds.cern.ch/record/2869272",
    year           = "2023",
}

@Report{ATLAS-TDR-19,
    author         = "{ATLAS Collaboration}",
    title          = "{ATLAS Insertable B-Layer: Technical Design Report}",
    type           = "ATLAS-TDR-19; CERN-LHCC-2010-013",
    year           = "2010",
    url            = "https://cds.cern.ch/record/1291633",
    related        = "ATLAS-TDR-19-addm",
    relatedstring  = "Addendum:",
}

@Article{CMS-EXO-10-010,
    author         = "{CMS Collaboration}",
    title          = "{Search for Dijet Resonances in \(7\,\text{TeV}\) \(pp\) Collisions at CMS}",
    journal        = "Phys. Rev. Lett.",
    volume         = "105",
    year           = "2010",
    pages          = "211801",
    doi            = "10.1103/PhysRevLett.105.211801",
    reportNumber   = "CERN-PH-EP-2010-035",
    eprint         = "1010.0203",
    archivePrefix  = "arXiv",
    primaryClass   = "hep-ex",
}

@Article{CMS-EXO-11-015,
    author         = "{CMS Collaboration}",
    title          = "{Search for Resonances in the Dijet Mass Spectrum from \(7\,\text{TeV}\) \(pp\) Collisions at CMS}",
    journal        = "Phys. Lett. B",
    volume         = "704",
    year           = "2011",
    pages          = "123",
    doi            = "10.1016/j.physletb.2011.09.015",
    reportNumber   = "CERN-PH-EP-2011-119",
    eprint         = "1107.4771",
    archivePrefix  = "arXiv",
    primaryClass   = "hep-ex",
}

@Article{CMS-EXO-11-094,
    author         = "{CMS Collaboration}",
    title          = "{Search for narrow resonances and quantum black holes in inclusive and \(b\)-tagged dijet mass spectra from \(pp\) collisions at \(\sqrt{s} = 7\,\text{TeV}\)}",
    journal        = "JHEP",
    volume         = "01",
    year           = "2013",
    pages          = "013",
    doi            = "10.1007/JHEP01(2013)013",
    reportNumber   = "CERN-PH-EP-2012-277",
    eprint         = "1210.2387",
    archivePrefix  = "arXiv",
    primaryClass   = "hep-ex",
}

@Article{CMS-EXO-12-016,
    author         = "{CMS Collaboration}",
    title          = "{Search for narrow resonances using the dijet mass spectrum in \(pp\) collisions at \(\sqrt{s} = 8\,\text{TeV}\)}",
    journal        = "Phys. Rev. D",
    volume         = "87",
    year           = "2013",
    pages          = "114015",
    doi            = "10.1103/PhysRevD.87.114015",
    reportNumber   = "CERN-PH-EP-2013-015",
    eprint         = "1302.4794",
    archivePrefix  = "arXiv",
    primaryClass   = "hep-ex",
}

@Article{CMS-EXO-12-052,
    author         = "{CMS Collaboration}",
    title          = "{Search for pair-produced resonances decaying to jet pairs in proton--proton collisions at \(\sqrt{s} = 8\,\text{TeV}\)}",
    journal        = "Phys. Lett. B",
    volume         = "747",
    year           = "2015",
    pages          = "98",
    doi            = "10.1016/j.physletb.2015.04.045",
    reportNumber   = "CERN-PH-EP-2014-293",
    eprint         = "1412.7706",
    archivePrefix  = "arXiv",
    primaryClass   = "hep-ex",
}

@Article{CMS-EXO-12-059,
    author         = "{CMS Collaboration}",
    title          = "{Search for resonances and quantum black holes using dijet mass spectra in proton--proton collisions at \(\sqrt{s} = 8\,\text{TeV}\)}",
    journal        = "Phys. Rev. D",
    volume         = "91",
    year           = "2015",
    pages          = "052009",
    doi            = "10.1103/PhysRevD.91.052009",
    reportNumber   = "CERN-PH-EP-2014-300",
    eprint         = "1501.04198",
    archivePrefix  = "arXiv",
    primaryClass   = "hep-ex",
}

@Article{CMS-JME-13-004,
    author         = "{CMS Collaboration}",
    title          = "{Jet energy scale and resolution in the CMS experiment in \(pp\) collisions at \(8\,\text{TeV}\)}",
    journal        = "JINST",
    volume         = "12",
    year           = "2017",
    pages          = "P02014",
    doi            = "10.1088/1748-0221/12/02/P02014",
    reportNumber   = "CERN-PH-EP-2015-305",
    eprint         = "1607.03663",
    archivePrefix  = "arXiv",
    primaryClass   = "hep-ex",
}

@Article{CMS-EXO-14-005,
    author         = "{CMS Collaboration}",
    title          = "{Search for Narrow Resonances in Dijet Final States at \(\sqrt{s} = 8\,\text{TeV}\) with the Novel CMS Technique of Data Scouting}",
    journal        = "Phys. Rev. Lett.",
    volume         = "117",
    year           = "2016",
    pages          = "031802",
    doi            = "10.1103/PhysRevLett.117.031802",
    reportNumber   = "CERN-EP-2016-090",
    eprint         = "1604.08907",
    archivePrefix  = "arXiv",
    primaryClass   = "hep-ex",
}

@Article{CMS-EXO-15-001,
    author         = "{CMS Collaboration}",
    title          = "{Search for narrow resonances decaying to dijets in proton--proton collisions at \(\sqrt{s} = 13\,\text{TeV}\)}",
    journal        = "Phys. Rev. Lett.",
    volume         = "116",
    year           = "2016",
    pages          = "071801",
    doi            = "10.1103/PhysRevLett.116.071801",
    reportNumber   = "CERN-PH-EP-2015-317",
    eprint         = "1512.01224",
    archivePrefix  = "arXiv",
    primaryClass   = "hep-ex",
}

@Article{CMS-EXO-16-032,
    author         = "{CMS Collaboration}",
    title          = "{Search for dijet resonances in proton--proton collisions at \(\sqrt{s} = 13\,\text{TeV}\) and constraints on dark matter and other models}",
    journal        = "Phys. Lett. B",
    volume         = "769",
    year           = "2017",
    pages          = "520",
    doi            = "10.1016/j.physletb.2017.02.012",
    reportNumber   = "CERN-EP-2016-277",
    eprint         = "1611.03568",
    archivePrefix  = "arXiv",
    primaryClass   = "hep-ex",
}

@Article{CMS-EXO-16-056,
    author         = "{CMS Collaboration}",
    title          = "{Search for narrow and broad dijet resonances in proton--proton collisions at \(\sqrt{s} = 13\,\text{TeV}\) and constraints on dark matter mediators and other new particles}",
    journal        = "JHEP",
    volume         = "08",
    year           = "2018",
    pages          = "130",
    doi            = "10.1007/JHEP08(2018)130",
    reportNumber   = "CERN-EP-2018-123",
    eprint         = "1806.00843",
    archivePrefix  = "arXiv",
    primaryClass   = "hep-ex",
}

@Article{CMS-EXO-17-021,
    author         = "{CMS Collaboration}",
    title          = "{Search for pair-produced resonances decaying to quark pairs in proton--proton collisions at \(\sqrt{s} = 13\,\text{TeV}\)}",
    journal        = "Phys. Rev. D",
    volume         = "98",
    year           = "2018",
    pages          = "112014",
    doi            = "10.1103/PhysRevD.98.112014",
    reportNumber   = "CERN-EP-2018-211",
    eprint         = "1808.03124",
    archivePrefix  = "arXiv",
    primaryClass   = "hep-ex",
}

@Article{CMS-EXO-19-012,
    author         = "{CMS Collaboration}",
    title          = "{Search for high mass dijet resonances with a new background prediction method in proton--proton collisions at \(\sqrt{s} = 13\,\text{TeV}\)}",
    journal        = "JHEP",
    volume         = "05",
    year           = "2020",
    pages          = "033",
    doi            = "10.1007/JHEP05(2020)033",
    reportNumber   = "CERN-EP-2019-222",
    eprint         = "1911.03947",
    archivePrefix  = "arXiv",
    primaryClass   = "hep-ex",
}

@Article{CMS-EXO-21-010,
    author         = "{CMS Collaboration}",
    title          = "{Search for resonant and nonresonant production of pairs of dijet resonances in proton--proton collisions at \(\sqrt{s} = 13\,\text{TeV}\)}",
    journal        = "JHEP",
    volume         = "07",
    year           = "2022",
    pages          = "161",
    doi            = "10.1007/JHEP07(2023)161",
    reportNumber   = "CERN-EP-2022-103",
    eprint         = "2206.09997",
    archivePrefix  = "arXiv",
    primaryClass   = "hep-ex",
}

@Booklet{ATL-PHYS-PUB-2010-013,
    author         = "{ATLAS Collaboration}",
    title          = "{The simulation principle and performance of the ATLAS fast calorimeter simulation FastCaloSim}",
    howpublished   = "{ATL-PHYS-PUB-2010-013}",
    url            = "https://cds.cern.ch/record/1300517",
    year           = "2010",
}

@Booklet{ATL-PHYS-PUB-2014-021,
    author         = "{ATLAS Collaboration}",
    title          = "{ATLAS Pythia~8 tunes to \(7~\text{TeV}\) data}",
    howpublished   = "{ATL-PHYS-PUB-2014-021}",
    url            = "https://cds.cern.ch/record/1966419",
    year           = "2014",
}

@Booklet{ATL-PHYS-PUB-2016-017,
    author         = "{ATLAS Collaboration}",
    title          = "{The Pythia~8 A3 tune description of ATLAS minimum bias and inelastic measurements incorporating the Donnachie--Landshoff diffractive model}",
    howpublished   = "{ATL-PHYS-PUB-2016-017}",
    url            = "https://cds.cern.ch/record/2206965",
    year           = "2016",
}

@Booklet{ATL-PHYS-PUB-2019-017,
    author         = "{ATLAS Collaboration}",
    title          = "{Multijet simulation for \(13\,\text{TeV}\) ATLAS Analyses}",
    howpublished   = "{ATL-PHYS-PUB-2019-017}",
    url            = "https://cds.cern.ch/record/2672252",
    year           = "2019",
}
 
\clearpage
 
\begin{flushleft}
\hypersetup{urlcolor=black}
{\Large The ATLAS Collaboration}

\bigskip

\AtlasOrcid[0000-0002-6665-4934]{G.~Aad}$^\textrm{\scriptsize 102}$,
\AtlasOrcid[0000-0002-5888-2734]{B.~Abbott}$^\textrm{\scriptsize 120}$,
\AtlasOrcid[0000-0002-1002-1652]{K.~Abeling}$^\textrm{\scriptsize 55}$,
\AtlasOrcid[0000-0001-5763-2760]{N.J.~Abicht}$^\textrm{\scriptsize 49}$,
\AtlasOrcid[0000-0002-8496-9294]{S.H.~Abidi}$^\textrm{\scriptsize 29}$,
\AtlasOrcid[0000-0002-9987-2292]{A.~Aboulhorma}$^\textrm{\scriptsize 35e}$,
\AtlasOrcid[0000-0001-5329-6640]{H.~Abramowicz}$^\textrm{\scriptsize 151}$,
\AtlasOrcid[0000-0002-1599-2896]{H.~Abreu}$^\textrm{\scriptsize 150}$,
\AtlasOrcid[0000-0003-0403-3697]{Y.~Abulaiti}$^\textrm{\scriptsize 117}$,
\AtlasOrcid[0000-0002-8588-9157]{B.S.~Acharya}$^\textrm{\scriptsize 69a,69b,m}$
% ...
\bigskip
\\

% ...
$^{29}$Physics Department, Brookhaven National Laboratory, Upton NY; United States of America.\\
% ...
$^{35}$$^{(a)}$Facult\'e des Sciences Ain Chock, R\'eseau Universitaire de Physique des Hautes Energies - Universit\'e Hassan II, Casablanca;$^{(b)}$Facult\'{e} des Sciences, Universit\'{e} Ibn-Tofail, K\'{e}nitra;$^{(c)}$Facult\'e des Sciences Semlalia, Universit\'e Cadi Ayyad, LPHEA-Marrakech;$^{(d)}$LPMR, Facult\'e des Sciences, Universit\'e Mohamed Premier, Oujda;$^{(e)}$Facult\'e des sciences, Universit\'e Mohammed V, Rabat;$^{(f)}$Institute of Applied Physics, Mohammed VI Polytechnic University, Ben Guerir; Morocco.\\
% ...
$^{49}$Fakult\"{a}t Physik , Technische Universit{\"a}t Dortmund, Dortmund; Germany.\\
% ...
$^{55}$II. Physikalisches Institut, Georg-August-Universit\"{a}t G\"ottingen, G\"ottingen; Germany.\\
% ...
$^{69}$$^{(a)}$INFN Gruppo Collegato di Udine, Sezione di Trieste, Udine;$^{(b)}$ICTP, Trieste;$^{(c)}$Dipartimento Politecnico di Ingegneria e Architettura, Universit\`a di Udine, Udine; Italy.\\
% ...
$^{102}$CPPM, Aix-Marseille Universit\'e, CNRS/IN2P3, Marseille; France.\\
% ...
$^{117}$Department of Physics, New York University, New York NY; United States of America.\\
% ...
$^{120}$Homer L. Dodge Department of Physics and Astronomy, University of Oklahoma, Norman OK; United States of America.\\
% ...
$^{150}$Department of Physics, Technion, Israel Institute of Technology, Haifa; Israel.\\
$^{151}$Raymond and Beverly Sackler School of Physics and Astronomy, Tel Aviv University, Tel Aviv; Israel.\\
% ...
$^{a}$ Also Affiliated with an institute covered by a cooperation agreement with CERN.\\
$^{b}$ Also at An-Najah National University, Nablus; Palestine.\\
$^{c}$ Also at Borough of Manhattan Community College, City University of New York, New York NY; United States of America.\\
$^{d}$ Also at Center for High Energy Physics, Peking University; China.\\
$^{e}$ Also at Center for Interdisciplinary Research and Innovation (CIRI-AUTH), Thessaloniki; Greece.\\
$^{f}$ Also at Centro Studi e Ricerche Enrico Fermi; Italy.\\
% ...
$^{m}$ Also at Department of Physics, King's College London, London; United Kingdom.\\
% ...

\end{flushleft}


\end{document}